\documentclass{article}
\usepackage{amssymb,amsmath,bm,hyperref,natbib,epsfig,fullpage}
\newcommand{\E}{\ensuremath{\operatorname{\mathbb{E}}}}
\newcommand{\matern}{Mat\'ern }
\begin{document}
\title{Lagrangian Time Series Models for Ocean Surface Drifter Trajectories}
\author{Adam M. Sykulski, Sofia C. Olhede, Jonathan M. Lilly and Eric Danioux}
\date{}
\maketitle
\begin{abstract}
This paper proposes stochastic models for the analysis of ocean surface trajectories obtained from freely-drifting satellite-tracked instruments. The proposed time series models are used to summarise large multivariate datasets and infer important physical parameters of inertial oscillations and other ocean processes. Nonstationary time series methods are employed to account for the spatiotemporal variability of each trajectory. Because the datasets are large, we construct computationally efficient methods through the use of frequency-domain modelling and estimation, with the data expressed as complex-valued time series. We detail how practical issues related to sampling and model misspecification may be addressed using semi-parametric techniques for time series, and we demonstrate the effectiveness of our stochastic models through application to both real-world data and to numerical model output.
\let\thefootnote\relax\footnote{Adam M. Sykulski and Sofia C. Olhede are with the Department of Statistical Science, University College London, Gower Street, London WC1 E6BT, UK (a.sykulski@ucl.ac.uk, s.olhede@ucl.ac.uk), tel: +44 20 7679 1872, fax: +44 20 3108 3105.}
\let\thefootnote\relax\footnote{Jonathan M. Lilly is with Northwest Research Associates, PO Box 3027, Bellevue, WA 98009-3027, USA (lilly@nwra.com), tel: +1 425 556 9055, fax: +1 425 556 9099.}
\let\thefootnote\relax\footnote{Eric Danioux is with the School of Mathematics, University of Edinburgh, Old College, South Bridge, Edinburgh EH8 9YL, UK (Eric.Danioux@ed.ac.uk), tel: +44 131 650 5060, fax: +44 131 650 6553.}
\end{abstract}
\section{Introduction}\label{S:Introduction}
Capturing spatial and temporal structure in high-dimensional datasets is an important theme in modern statistics, with substantial methodological challenges, see for example recent work by \cite{davis2013statistical}, \cite{fuentes2013nonparametric} and \cite{guinness2013interpolation}.  Examples of such endeavours include environmental studies \citep{bowman2009spatiotemporal}, brain imaging \citep{erhardt2012simtb}, and climate modelling \citep{steinhaeuser2012multivariate}.  In this paper we develop methods for {\em Lagrangian time series}, an important type of spatiotemporal data which results from tracking the spatial movement of objects over time.  Such time series are commonly encountered in oceanography \citep{griffa2007lagrangian}, wildlife tracking \citep{schofield2007novel}, and wireless sensing of traffic flow \citep{herrera2010incorporation}.

In oceanography, freely-drifting, or Lagrangian, instruments are one of the primary ways of directly observing the ocean currents.  For studying the currents at a single vertical level, far greater and more cost-effective spatial coverage can be obtained from freely-drifting instruments than from fixed-location instruments.  The position readings from Lagrangian instruments approximate the motion of a fluid particle over time.  Complementary information is obtain from fixed-location or {\em Eulerian} time series, which in oceanography typically involve records of currents and water properties at a range of depths from moored instruments.  Both types of time series contribute to our understanding of the ocean circulation, but the statistical treatment of Lagrangian time series is perhaps more subtle and challenging due to the spatial {\em and} temporal considerations that must be made for each individual time series \citep[see e.g.][]{griffa2007lagrangian}.  Typically, Lagrangian time series are highly nonstationary, as their statistics evolve in time.

Oceanographic Lagrangian instruments may be classed into two main types.  {\em Surface drifters} follow the water motion at the ocean's surface, and are tracked by satellite, see \cite{lumpkin07}.  Subsurface {\em Lagrangian floats} are neutrally buoyant instruments that drift with the currents at some deeper depth, and are tracked acoustically while underwater, see \cite{rossby07-lapcod}.  This paper deals primarily with surface drifter trajectories, which differ from those of deep floats through the presence of strong {\em inertial oscillations}, a characteristic feature of the ocean surface layer that arises from the interaction of wind forcing with the Coriolis force \cite[see e.g.][]{pollard1970comparison}.  However we expect the modelling approach taken in this paper will also prove useful for float trajectories.

\begin{figure}[h]
\centering
\includegraphics[width=0.56\textwidth]{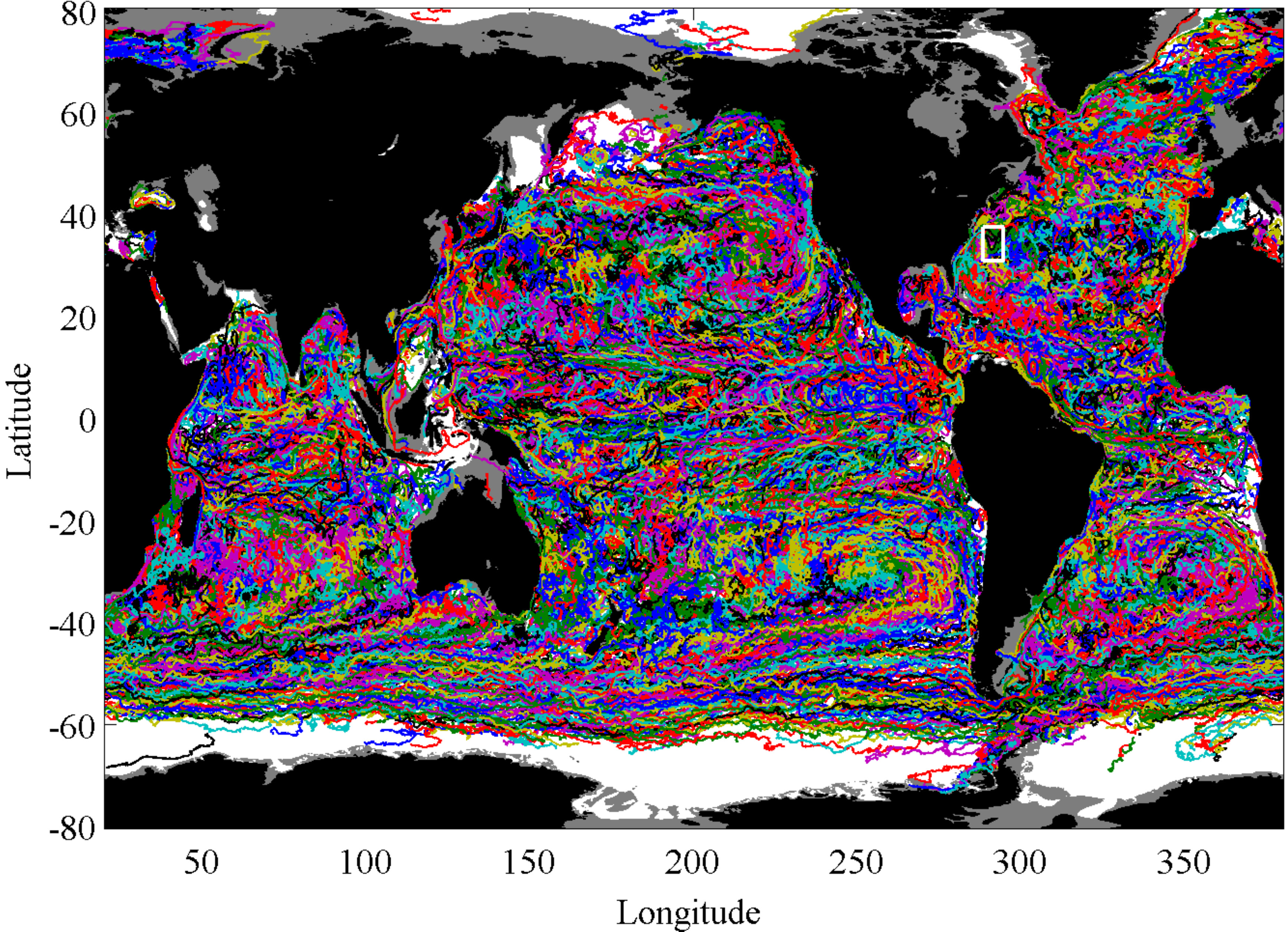}
\includegraphics[width=0.41\textwidth]{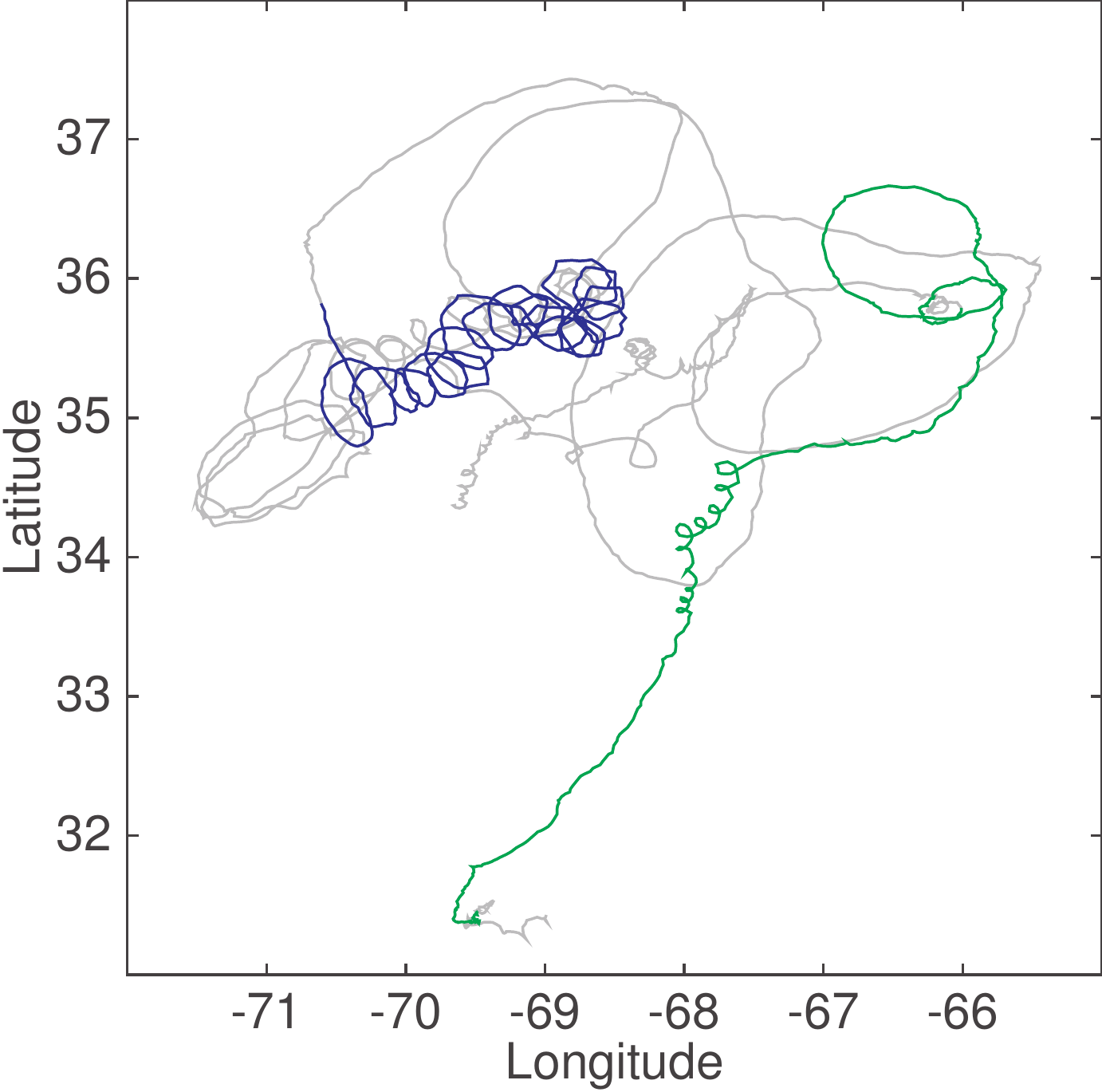}
\caption{\label{FigIntro}Trajectories from the Global Drifter Program (left) and a 250-day trajectory of a North Atlantic drifter (right).  In the left panel, each colour corresponds to a different drifter, and the location of the trajectory shown at right is indicated by a small white box.  The continents are shown as black shading and the continental shelf, with a depth of 500 metres, as grey shading.  In the right panel, two 50-day portions that are analysed in Figure~\ref{FigIntro2} are highlighted in blue and green respectively.
}
\end{figure}

Figure~\ref{FigIntro} (left) displays the global array of surface drifter trajectories from the National Oceanic and Atmospheric Administration's (NOAA) Global Drifter Program (GDP, www.aoml.noaa.gov/phod/dac).  In total over 10,000 drifters have been deployed since 1979, representing nearly 30 million data points of positions at 6-hour intervals.  Figure~\ref{FigIntro} (right) displays a portion of one North Atlantic drifter trajectory, Global Drifter Program ID \#44000, which exhibits clear nonstationary structure.  The analysis of this large heterogenous dataset is a major source of information regarding ocean circulation, an important component of the global climate system, see e.g. \cite{lumpkin07}.

In the oceanographic literature, the analysis of Lagrangian data is primarily restricted to nonparametric estimates of the first-order and second-order moments of the velocity time series \cite[e.g.][]{klocker2012reconciling,lacasce2008statistics}.  Nonstationarity is typically dealt with by isolating analysis to localised time periods and spatial regions.  In this paper we take an alternative approach, and seek to build statistical models using knowledge of the anticipated physical structure in the data.  To this end, we construct physically-motivated stochastic models, utilising oceanographic knowledge as well as empirical evidence from data, and in particular building on the work of \cite{griffa2007lagrangian}, \cite{lacasce2008statistics}, \cite{pollard1970comparison}, and \cite{rupolo1996lagrangian}.  Our models respect {\em multiscale structure}---the superposition of important features at different temporal scales---and accommodate nonstationarity by allowing for time dependence arising from the heterogeneity of the oceans.  To conveniently represent circular oscillatory motion, we express the data as complex-valued time series, which is simpler to the alternative of {\em coupling} bivariate real-valued time series (as in~\cite{veneziani2004oceanic}), as each trajectory is then represented in one time series sample.

\begin{figure}[h]
\includegraphics[width=0.49\textwidth]{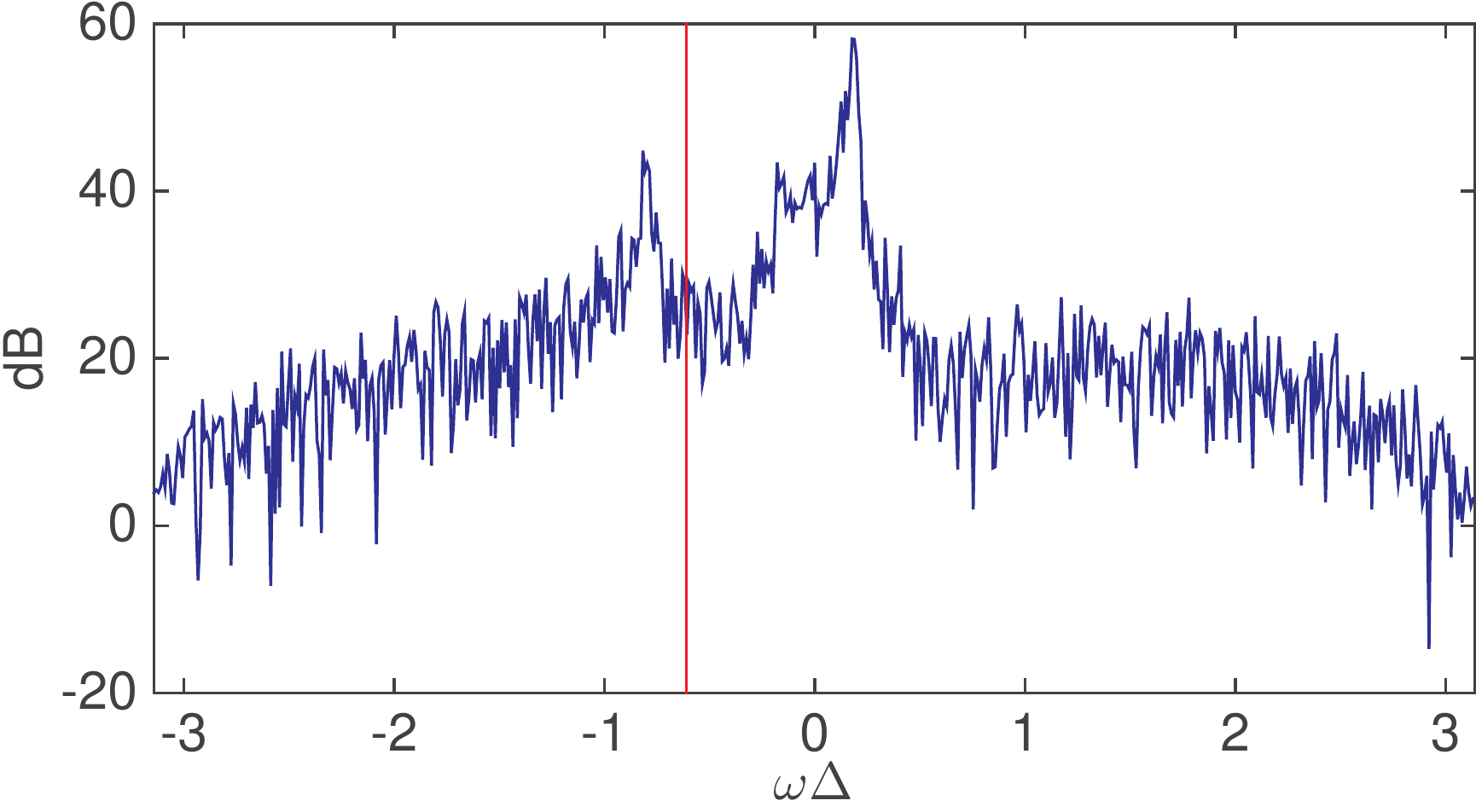}
\includegraphics[width=0.49\textwidth]{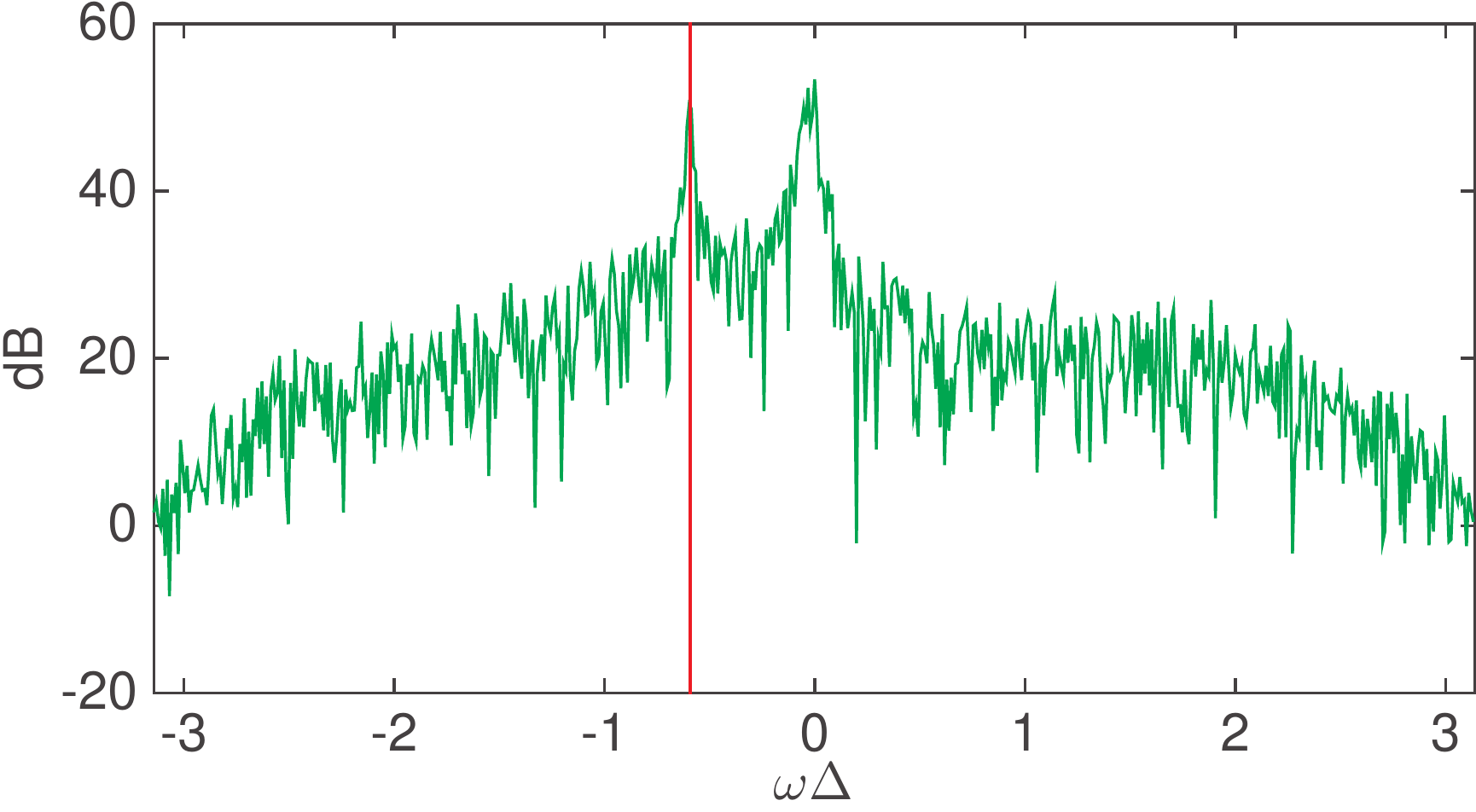}
\vspace{-3mm}
\caption{\label{FigIntro2}Spectral estimates of the blue (left) and green (right) portions of the velocity time series of the drifter displayed in Figure~\ref{FigIntro} (right), where the red lines denote the Coriolis frequency.} 
\end{figure}

As examples, in Figure~\ref{FigIntro2} we display spectral estimates of the complex-valued velocity time series corresponding to the trajectory displayed in the right panel of Figure~\ref{FigIntro}.  Here positive and negative frequencies represent anti-clockwise or clockwise oscillations, respectively, and the spectral energy is given on a decibel scale.  Spectra are displayed from two different portions of the time series, as indicated by the colours. Both spectra display clear multiscale structure with a pronounced peak near zero frequency, reflecting the {\em turbulent background flow}, and a negative-frequency peak capturing inertial oscillations in the vicinity of the local Coriolis frequency, defined subsequently. The amplitudes and positions of the peaks differ markedly between the two time periods, indicating nonstationarity, since a stationary process would by definition have the same expected spectrum at all times. New statistical summaries that capture such variability are expected to prove useful to oceanographers, particularly in important applications such as climate forecasting.

In this paper we show how one may consistently capture ocean variability from drifter trajectories, by aggregating two simple stochastic processes.  This model allows parameters to be estimated locally in time using short windows, yielding parameter estimates with high temporal and spatial resolution.  The use of maximum likelihood theory quantifies the uncertainty of these estimates.  While the use of stochastic processes to study Lagrangian times series is common in oceanography---see for example \cite{berloff2002material}, \cite{veneziani2004oceanic}, and \cite{lacasce2008statistics}---the stochastic model presented here is the first to explicitly include the effects of inertial oscillations.  We show how these oscillations may be accurately captured using the complex-valued Ornstein-Uhlenbeck (OU) process \citep{jeffreys1940variation,Arato1962estimation}.  Furthermore, we find that existing stochastic models for the turbulent background process are insufficiently flexible to capture key features of observed spectra.  In particular, by limiting spectral slopes to even integral values, these models implicitly assume Markovian structure, which observed data does not support.  Instead we show how models for the background can be generalised and unified using the {\em \matern process} \citep{stein1999interpolation}, a stationary fractional stochastic process.  This process has been mainly used for spatial models, whereas its usage for temporally sampled processes remains largely unexplored.

Applying our models to real-world data requires several methodological refinements. Specifically, we show how to account for temporal sampling effects such as aliasing and leakage, and how to employ semi-parametric fitting procedures both in frequency and in time in order to account for nonstationarity and model misspecification.  This latter task involves unifying the ideas of  \cite{robinson1995gaussian} on semi-parametric fitting of spectral slopes, with those of \cite{van2006semiparametric} and \cite{robinson89} on semi-parametric modelling of time-varying parameters, which are modified herein to accommodate complex-valued time series.  Finally, we develop several variants of our model together with the statistical procedures for selecting between them, allowing us to test important physical hypotheses.  In particular, we wish to examine whether currents due to {\em eddies} (persistent rings of swirling currents) may act to shift the location of inertial oscillations away from the local Coriolis frequency, as has long been expected on theoretical grounds \citep{kunze85-jpo}, but only rarely observed \citep{elipot10-jgr}.

The paper is organised as follows.  In Section~\ref{S:Model} we describe our model and its relation to the different physical processes.  In Section~\ref{S:Estimation}, we detail the semi-parametric estimation procedure, accounting for sampling effects, spatiotemporal variability and model misspecification.  We then perform an extensive analysis of our model in Section~\ref{S:Simulations} using both real-world data and numerical model output.  Section~\ref{S:Conclusion} presents conclusions and directions for future work.
\section{Stochastic models for surface drifter data}\label{S:Model}
In this section we will describe in detail the two key physical ocean processes we model: the {\em inertial oscillation} and the {\em turbulent background}.  We note that eddies are not modelled herein, as they are intermittent phenomena that are absent from many time series; these may be analysed using previously developed time-frequency methodology \citep{lilly11-grl}.

We require stochastic models that are physically motivated and that accurately describe the data we observe.  We model the drifter velocities as a superposition of Gaussian processes, and therefore the aggregated model describes the data to the second order (i.e.  in terms of variances and covariances).  Higher-order models are not desirable for this application as they introduce additional parameters and complicate estimation procedures, particularly as longer time windows will be required to fit the model, decreasing the temporal and spatial resolution of any summaries from the data.  Furthermore, drifter data supports the assumption of Gaussian distributed velocities \cite[see][]{lacasce2008statistics}.

We first formulate the continuous-time representation for the inertial oscillation as a damped stochastic oscillator, corresponding to the complex-valued Ornstein-Uhlenbeck (OU) stochastic process.  The turbulent background process is then modelled using the \matern process \citep{stein1999interpolation} which flexibly accounts for the various spectral  slopes observed in the data \citep{rupolo1996lagrangian,sanderson1991fractal}.  We also provide theoretical reasoning for {\em not} modelling the background as a pure power law process such as fractional Brownian motion or fBm \citep{mandelbrot1968fractional}, as is often suggested in the oceanographic literature \citep{lacasce2008statistics}.  The aggregated stochastic model is then formalised by combining these two components.
\subsection{Inertial oscillation} \label{SS:Inertial}
An inertial oscillation is a phenomena attributed to the Coriolis force that arises from the Earth's rotation.  The interaction of wind forcing with the Coriolis force generates circular oscillations of water parcels, leading a peak in the velocity spectrum at the Coriolis or inertial frequency, which varies as a function of latitude.  The inertial peak occurs at negative frequencies for the anticlockwise oscillations occurring in the Northern Hemisphere, and at positive frequencies for the clockwise oscillations in the Southern Hemisphere---as evidenced by the negative frequency peak in Figure~\ref{FigIntro2} for a North Atlantic drifter.

A well-known deterministic model for inertial oscillations is given as a set of coupled ordinary differential equations by \cite{pollard1970comparison}, see also \cite{early2012forces}, in which the velocities of a water parcel are modelled by
\begin{equation}
\frac{\partial u}{\partial t}+f_o\,v  =  F-cu, \quad\quad
\frac{\partial v}{\partial t}-f_o\,u  =  G-cv,
\label{e:pollard}
\end{equation}
where $u$ and $v$ respectively correspond to the eastward and northward velocities, $c>0$ is an unknown linear damping term, and $F$ and $G$ are time-varying forcing functions associated with surface winds. The parameter $f_o$ is the  Coriolis frequency given by
\begin{equation}
f_o=-2\Psi\mathrm{sin}(\varphi),
\label{e:CorrFreq}
\end{equation}
where $\Psi=7.29\times10^{-5}$ rad/s is the rotation rate of the Earth, and $\varphi$ is the latitude in radians.  Note that we depart from convention by defining $f_o$ to be {\em negative} in the Northern Hemisphere, in order to reflect the rotation direction of the inertial oscillations.  This means that the terms $f_o \, v$ and $f_o\, u$ in equation~\eqref{e:pollard} appear with the reverse of their usual signs.

As noted by \cite{kunze85-jpo}, the effect of eddies can shift the oscillation frequency from the Coriolis frequency to a nearby frequency.  Specifically, we would expect to see the inertial oscillations shifted to $f=f_o-\omega_e$, where $\omega_e$ is the angular frequency of the eddy. From dynamical limits of eddies (see~\cite{thomas07-aha}), a reasonable range for $f$ is anywhere between $0.5f_o$ and $1.5f_o$.

Motivated by above results, we model an inertial oscillation as the stochastic process that is analogous to the deterministic oscillator of equation~\eqref{e:pollard},  but with the effective inertial frequency as a free parameter, to allow for shifts due to eddies.  The complex-valued velocity, $z(t)=u(t)+iv(t)$, is then governed by the stochastic differential equation (SDE)
\begin{equation}
dz(t)=(i\omega_o-c)z(t)dt +A dQ(t),
\label{e:complexou}
\end{equation}
where $dQ(t)$ are independent complex-valued Brownian increments (see \cite{mandelbrot1968fractional}) which represent an idealised stochastic model for the wind forcing, combining the effects of $F$ and $G$ from equation~\eqref{e:pollard}.  Equation~\eqref{e:complexou} is recognised as the complex Ornstein-Uhlenbeck (OU) process \citep{jeffreys1940variation,Arato1962estimation}.  The parameter $\omega_o$ dictates the frequency of the inertial oscillation in radians per unit time, and may not necessarily equal the local Coriolis frequency $f_o$.  The damping parameter $c>0$ ensures the process is mean-reverting, in other words, the process is guaranteed to eventually return to its expected value ($z=0$, in this case) from any initial condition.  Finally, $A>0$ sets the magnitude of the variability in this stochastic process, and reflects the strength of the surface wind stress forcing.  The complex OU process is Markovian, meaning that conditional on a present value, past and future values are independent.  The process is also stationary, so that its statistics are independent of time (and zero-mean) provided that initial conditions at time $t_o$ are set as $z(t_o)\sim\mathcal{N}(0,A^2/2|c|)$ where $A^2/2|c|$ is the process variance. For any other initial condition, the process converges quickly to a stationary process, because the effect of the initial condition will decay exponentially from the initial time.

Note that we use a complex-valued representation, as opposed to the bivariate coupled representation as in equation~\eqref{e:pollard}.  This allows for a clear separation of variability that oscillates clockwise or anticlockwise, which becomes very useful when performing semi-parametric inference, as we shall discuss in Section~\ref{S:Estimation}.  The stochastic model of equation~\eqref{e:complexou} assumes white noise forcing, i.e.  noise that is uncorrelated in time, whereas the full surface winds are known to generally be somewhat red \citep{gille05-jaot}, corresponding to a correlated random process.  The idealisation of the wind forcing as white is nevertheless reasonable for our purposes.  This is because the complex OU process has a nearly singular response at the effective Coriolis frequency, so that the wind forcing may usefully be regarded as {\em locally white} in this vicinity.  Moreover, the inertial oscillation model will only be used to capture spectral energy that is localised around the inertial peak; the rest of the spectrum will be captured by a model for the turbulent background, as described in Section~\ref{SS:Background}.

The autocovariance function and power spectral density of the continuous-time complex OU process, starting the SDE with stationary initial conditions discussed above, are found to be
\begin{eqnarray}
s^{(o)}(\tau)&=&\E\left\{z(t)z^*(t+\tau)\right\}=\frac{A^2}{2c}e^{i\omega_o \tau}e^{-c|\tau|},
\label{e:ouacvs} \\
S^{(o)}(\omega)&=&\int_{-\infty}^\infty s^{(o)}(\tau)e^{-i\omega\tau}d\tau=\frac{A^2}{(\omega-\omega_o)^2+c^2},
\label{e:ouspectrum}
\end{eqnarray}
where $\E$ is the expectation operator, and where $z^*(t)$ is the complex conjugate of $z(t)$. The superscript $(o)$ notation is used to denote that this is for the OU component of the overall model. Note that the autocovariance $s^{(o)}(\tau)$ decays exponentially, indicating that the complex OU process is a short-memory process.  The power spectral density $S^{(o)}(\omega)$ forms a Fourier pair with the autocovariance and decays in frequency proportionally to $|\omega-\omega_o|^2$ for $|\omega-\omega_o|\gg c$, where $c$ is the damping parameter.

A real-valued process is obtained when we set $\omega_o=0$  in equation~\eqref{e:complexou} and let $dQ(t)$ become real-valued.  This then describes the motion of a massive Brownian particle subject to friction, and is identical to a Continuous Auto-Regressive or CAR(1) process \citep{brockwell2007continuous}, the continuous-time analogue of a discrete-time stationary auto-regressive or AR(1) process.  This real-valued forced/damped stochastic system results in a stochastic process known as the Ornstein-Uhlenbeck (OU) process.  The extension of the real-valued OU process to complex values allows for oscillations which can be clockwise or anticlockwise, depending upon on the sign of $\omega_o$.  The complex-valued OU process was originally used to describe the Chandler Wobble \citep{jeffreys1940variation}, a periodic oscillation in the axis of the earth's rotation relative to the solid earth.  A related bivariate-coupled OU process has been proposed in the oceanographic literature for a different purpose, namely to model the effects of eddies in Lagrangian  trajectories \citep{veneziani2004oceanic}.
\subsection{Turbulent background}\label{SS:Background}
A second major component commonly observed in Lagrangian data is a red background process, reflecting the large-scale turbulent flow of the ocean currents \cite[e.g.][]{rhines79-arfm}.  Stochastic models have already been been proposed for the background in works such as \cite{sawford1991reynolds} and \cite{berloff2002material}; see \cite{lacasce2008statistics} for a review.  The proposed models  are all {\em integer order} in that they assume that the velocity (1st order), the acceleration (2nd order), or the hyper-acceleration (3rd order) is a Markovian process, such as Brownian motion or the Ornstein-Uhlenbeck (OU) process.  However, the evidence in the literature is that spectral slopes of Lagrangian velocities do not correspond to the power of an even integer \citep{rupolo1996lagrangian,sanderson1991fractal}.  This suggests the use of fractional, instead of Markovian, processes to model the background.

We therefore propose using the complex-valued \matern process to model background velocities.  The \matern process is a stationary fractional Gaussian process that essentially encompasses previously proposed integer-order models as special cases.  We use an isotropic complex-valued \matern process, analogous to \cite{gneiting2010matern} for spatial processes, corresponding to independent and identically distributed velocities in the real (eastward) and imaginary (northward) components.  This process is defined by either the autocovariance or power spectral density which are of the form  \cite[see][]{stein1999interpolation}
\begin{eqnarray}
s^{(m)}(\tau)&=&\frac{B^2}{2^{\alpha-1/2}\pi^{1/2}\Gamma(\alpha)h^{2\alpha-1}}(h|\tau|)^{\alpha-1/2}\mathcal{K}_{\alpha-1/2}(h|\tau|),\label{e:maternacvs}\\
S^{(m)}(\omega)&=&\frac{B^2}{(\omega^2+h^2)^{\alpha}},
\label{e:maternspectrum}
\end{eqnarray}
where $\Gamma(\alpha)$ is the Gamma function, and $\mathcal{K}_\eta$ is the modified Bessel function of the second kind of order $\eta$ \citep[][p.~374]{abramowitz1964handbook}. The superscript $(m)$ notation is used to denote the \matern component of the overall model. The damping parameter, $h>0$, is strictly greater than zero, while the slope parameter, $\alpha>1/2$, controls the degree of smoothness or differentiability of the process \cite[][]{gneiting2012estimators}.

The \matern process generalises and unifies the various stochastic models for the turbulent background proposed in the oceanographic literature.  Specifically, using positive even-integer values of $\alpha$ closely approximates the integer-order stochastic models reviewed in \cite{lacasce2008statistics}.  Furthermore, the \matern process can also model the non-integer slopes of 3.3 observed by \cite{rupolo1996lagrangian}, and the 2.5 implied by the fractal dimension analysis of \cite{sanderson1991fractal}---corresponding to $\alpha=1.65$ and $\alpha=1.25$, respectively---both of which require a fractional stochastic process.  Therefore the \matern process provides a flexible model that can accommodate a range of possible behaviours of the background process.  This allows one to formalise, for example, statistical tests for the hypothesis of Markovian velocities or accelerations, a useful application of the methods developed here that will be addressed in the future.  

A significant feature of the \matern process is its behaviour at low frequencies.  Previous works such as \cite{rupolo1996lagrangian} and \cite{sanderson1991fractal} appear to propose the use of a Pure Power Law process \citep[e.g.][]{percival2006wavelet} to model the background.  As the name implies, such processes have spectral energy that decays exactly proportionally to some power of the frequency at all frequencies.  This would correspond to setting $h=0$ in equation~\eqref{e:maternspectrum}.  The resulting spectrum is no longer that  of a \matern process, but instead yields the spectrum of fractional Brownian motion (fBm) for the range $1/2<\alpha<3/2$, or standard Brownian motion for $\alpha=1$.  Fractional Brownian motion is a self-similar Gaussian process~\citep{beran1994statistics} which is nonstationary and has variance growing proportionally to $t^{2\alpha-1}$ \citep{mandelbrot1968fractional}.  This process is only mean-reverting when $\alpha<1$.  While the idea of a pure power law process is appealing, the implied nonstationarity makes this unsuitable for modelling drifter velocities.  As time increases, velocities generated from such a model will tend to increase without bound, eventually becoming unphysically large, a consequence of the singularity of fBm at zero frequency. 

The \matern process thus has the desirable features of fBm without its drawbacks.  For $\omega\gg h$, the process exhibits power law decay, whereas for $h\rightarrow\infty$, or equivalently for $\omega\rightarrow0$, the process behaves like a white noise process, as could be expected if we were to observe Lagrangian velocities sparsely over long very time intervals.  The \matern process provides a continuum between these two regimes over different timescales, controlled by the value of $h$, which has the units of an inverse timescale parameter.  The \matern process is stationary and {\em not} strictly self-similar, so velocities generated from this process will not drift to large values over long timescales.  Yet the \matern process still has the same desirable mid-to-high frequency behaviour of fBm that is observed in the data over shorter timescales.

\begin{figure}[h]
\centering
\includegraphics[width=0.95\textwidth]{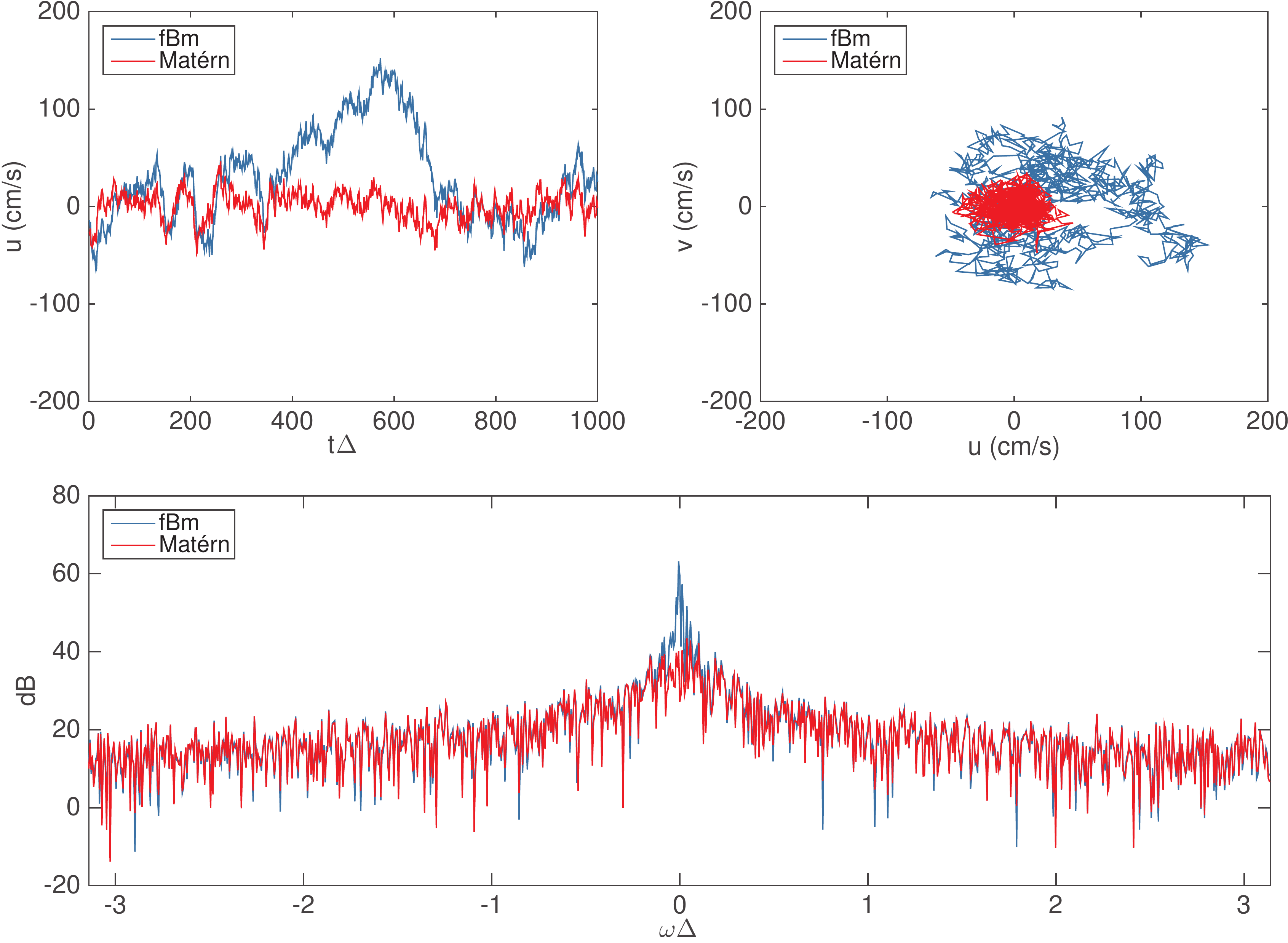}\vspace{-3mm}
\caption{Comparison of a \matern process with fractional Brownian motion (fBm). The top left panel displays the real-valued (eastward) component of a 1000-point complex-valued \matern process and a complex-valued fBm process, generated with the values $B=10$ and $\alpha=0.9$, and with $h=0.1$ for the \matern process.  The top right panel shows the complex-valued trajectories, with the eastward and northward processes being independent and identically distributed.  The bottom panel displays the periodograms of both complex-valued time series on a decibel scale.}
\label{fig:fbm}
\end{figure}

To illustrate the differences between the \matern process and fBm, in Figure~\ref{fig:fbm} we show sampled simulated time series and corresponding spectra for a \matern and fBm process having identical amplitude and decay parameters. The parameter values are based of typical maximum likelihood fits to drifter trajectories. We overlay the spectra for fBm and the \matern processes to show their similarity across most frequencies, but notice their markedly different generated time series.  fBm has a tendency to drift to very large values ($>100$ cm/s), which is caused by the low-frequency energy, whereas the \matern process generates more realistic drifter velocities. The mid-to-high frequency behaviour, on the other hand, is indistinguishable between the two processes.
\vspace{-1mm}
\subsection{The aggregated model}\label{SS:Complete}
We now combine the two models for the inertial oscillation and turbulent background process to form an aggregated statistical model for drifter velocity time series, characterised by the spectrum
\begin{equation}
S(\omega;\bm{\theta})= \frac{A^2}{(\omega-\omega_o)^2+c^2}+\frac{B^2}{(\omega^2+h^2)^\alpha},
\label{eq:model1}
\end{equation}
where $\bm{\theta}=(A,B,\omega_o,c,h,\alpha)$ is an array of model parameters.  $S(\omega;\bm{\theta})$ is therefore a 6-parameter model: $\omega_o$, $c>0$, and $A>0$ respectively correspond to the frequency, damping and amplitude of the inertial oscillation, and $B>0$, $\alpha>1/2$, and $h>0$ respectively correspond to the amplitude, smoothness and timescale parameter of the turbulent background process.  We use this model to estimate parameters from observed time series, providing useful summaries of structure as well as a physical interpretation of the processes that occur at the ocean surface.

Several useful special cases may be obtained by considering various model parameters to be fixed.  Fixing the frequency of the inertial oscillation at the local Coriolis frequency $\omega_o=f_o$ reduces our model to 5 unknown parameters, and comparison of the five-parameter and six-parameter models then allow us to assess whether sufficient statistical information is present to determine that a frequency shift has occurred.   Alternatively, setting $h=0$ models the background as fBm, which can then compared with the \matern process to determine whether the inverse timescale parameter $h$ can be resolved from short time series.  In addition, we can also create simpler 2 or 3 parameter models consisting only of background flow, when no inertial oscillation is expected to be present, by removing the complex-valued OU component from the model.  In Section~\ref{SS:Modelchoice} we show how one may select between these model variants using likelihood ratio tests.  Overall, our stochastic model is the first of its kind in terms of capturing ocean surface variability and structure for both the background and inertial oscillations. The model is sufficiently parsimonious such that summaries can be made from relatively short datasets, yet sufficiently rich to characterise the key variability of the drifter trajectories.  This is demonstrated in Section~\ref{S:Simulations} with real-world data and output from numerical models.
\section{Estimating parameters of the model}\label{S:Estimation}
In this section we first outline the estimation of the parameters of our stochastic model for any drifter time series, where we describe how to account for sampling effects explicitly.  Then we address how semi-parametric procedures are used in frequency to account for model misspecification and sampling artefacts, followed by a sample application.  We then show how nonstationarity of drifter time series may be modelled semi-parametrically, leading to the estimation of  time-varying parameters.  Finally, we explain how likelihood ratio tests can be used to select between variants of the stochastic model formalised in the preceding section.
\subsection{Frequency domain estimation using the Whittle likelihood}\label{SS:Whittle}
We estimate the parameters of our stochastic model in the frequency domain, as opposed to the time domain, for reasons to be discussed shortly.  To do this we require the periodogram, the modulus squared of the discrete Fourier transform of the regularly sampled time series $\{Z_t\} = Z_1,\ldots,Z_N$, defined as
\begin{equation}
\hat S_{Z}(\omega) = \frac{\Delta}{N}\left|\sum_{t=1}^{N}(Z_t-\bar{Z})e^{-it \omega \Delta}\right|^2,
\label{eq:periodogram}
\end{equation}
where the sample mean $\bar{Z}=\sum_{t=1}^N Z_t/N$ is subtracted, and where $\Delta$ is the sampling interval.  The periodogram is defined at the {\em Fourier frequencies}, $\omega = \pm\frac{2\pi}{N\Delta}(1,\ldots,\frac{1}{2}(N-1))$ when $N$ is odd and $\omega = \pm \frac{2\pi}{N\Delta} (1,\ldots,\frac{1}{2}N)$ when $N$ is even. The zero frequency is excluded here, as this information is lost by subtracting the mean.  As the data is complex-valued, the spectrum may be asymmetric about zero, and spectral estimates will therefore have $2N-2$ degrees of freedom, corresponding to  two degrees of freedom at each non-zero Fourier frequency.   The basic form of the frequency-domain likelihood is given by the {\em Whittle likelihood} \citep{dzaparidze1974new}
\begin{equation}
\ell_W(\bm{\theta})=  -\sum_{\omega\in\Omega}\left(\frac{\hat S_{Z}(\omega)}{S(\omega;\bm{\theta})}+\mathrm{log}\left(S(\omega;\bm{\theta})\right)\right),
\label{e:whittlefunc}
\end{equation}
where $\Omega$ denotes the set of frequencies chosen for comparing the modelled and estimated spectra.

It is well known however that the periodogram is a poor estimator for the spectral density \citep{thomson82-ieee,percival1993spectral}.  One reason for this is {\em broadband leakage}, a phenomena in which spectral energy spreads to distant parts of the spectrum.  Leakage is particularly problematic for data having spectra with a high dynamic range, such as a \matern process with a high $\alpha$ value, and can cause considerable bias in spectral estimates and subsequent parameter estimates.  To minimise these problems, \textit{tapered} spectral estimates are often used to reduce broadband leakage, at the expense of increased local bias and correlation between nearby Fourier frequencies; see \cite{percival1993spectral} for more details.  In addition, direct spectral estimates, including the periodogram as well as tapered estimates, are affected by {\em aliasing}.  In this phenomenon, all spectral energy higher than the Nyquist frequency $\omega=\pm \pi/\Delta$---the highest observable frequency---is wrapped back into the observable spectrum, at a Fourier coefficient that is a multiple of $2\pi/\Delta$ radians away.

The use of the periodogram for our purposes however is as an intermediate step in obtaining estimates of parameters for our parametric model.  As a result, we may address the known deficiencies of the periodogram by explicitly accounting for them during the estimation procedure.  We do this by fitting the periodogram against not the true spectrum, but rather with the {\em expected periodogram} that would be obtained from a true spectrum based on the model. This accounts for bias from both leakage and aliasing. Thus we follow the approach of \cite{sykulski2013whittle}, which introduces the {\em blurred} Whittle likelihood 
\begin{equation}
\ell_b(\bm{\theta})= -\sum_{\omega\in\Omega}\left(\frac{\hat S_Z(\omega)}{\bar{S}(\omega;\bm\theta)}+\mathrm{log}\left(\bar{S}(\omega;\bm\theta)\right)\right),
\label{e:whittlefunc2}
\end{equation}
where the expected periodogram, denoted $\bar{S}(\omega;\bm\theta)$, is conveniently given in terms of the true autocovariance function as
\begin{equation}
\bar{S}(\omega;\bm\theta)=
\Delta\sum_{\tau=-(N-1)}^{N-1}\left(1-\frac{|\tau|}{N}\right)s_\tau(\bm\theta) e^{-i\omega\tau\Delta}.
\label{eq:meanperiodogram}
\end{equation}
We then find the best fit parameters $\widehat{\bm{\theta}}^{(b)}$ by maximising the blurred likelihood function $\ell_b(\bm\theta)$.  In our model, the autocovariance function $s_\tau(\bm\theta)$ is the sum of the autocovariance sequences given in equations~\eqref{e:ouacvs} and \eqref{e:maternacvs}, leading to
\begin{equation}
s_\tau(\bm\theta)=\frac{A^2}{2c}e^{i\omega_o \tau}e^{-c|\tau|}+\frac{B^2\pi^{1/2}}{2^{\alpha-3/2}\Gamma(\alpha)h^{2\alpha-1}}(h|\tau|)^{\alpha-1/2}\mathcal{K}_{\alpha-1/2}(h|\tau|).
\end{equation}
Equation~\eqref{eq:meanperiodogram} makes use of the fact that the periodogram is a Fourier pair with the {\em biased} sample autocovariance sequence \citep{percival1993spectral}.  This form of likelihood is discussed more in \cite{sykulski2013whittle} where it is argued that in forming the Whittle likelihood for parametric spectral inference, the periodogram should be used rather than tapered spectral estimates.  This is so because we now simply wish to minimise errors arising from correlation between nearby frequencies, as the bias from leakage and aliasing has been explicitly accounted for.

We perform Whittle likelihood instead of classical time-domain maximum likelihood for two reasons.  Firstly, time-domain likelihood requires the inversion of a covariance matrix.  This is more computationally expensive---$\mathcal{O}(N^3)$ where $N$ is the length of the data---than the Whittle likelihood, which can be performed in $\mathcal{O}(N\mathrm{log}N)$ operations.  Secondly, with the Whittle likelihood we have the option to perform {\em semi-parametric} modelling and estimation {\em in the frequency domain,} by excluding certain Fourier frequencies from the set $\Omega$ in the likelihood summation.  There are a variety of reasons to do this, including dealing with model misspecification and interpolation, which we detail more in Section~\ref{SS:Windows}.  Such procedures are not easily performed in the time domain.  Alternatives to Whittle likelihood estimation in the frequency domain include using least squares techniques to estimate the parameters.  This technique has drawbacks, however, as it is not possible to take aliasing effects into account and furthermore, the estimates are known to be inefficient \citep{robinson1995gaussian}.

An additional useful feature of Whittle likelihood estimation is that we can easily compute estimates for the distributions of the parameter estimates $\widehat{\bm\theta}$ from a single time series using asymptotic theory \citep[][p.~104--109]{dzhaparidze1986parameter}
\begin{equation}
\sqrt{N}(\widehat{\bm\theta}-\bm\theta_o) \xrightarrow{d} \mathcal{N}(0,\mathcal{F}^{-1}),
\label{eq:fisher}
\end{equation}
where $\mathcal{F}$ is the Fisher information matrix and $\bm\theta_o$ are the ``true'' parameter values. The above equation states that $\sqrt{N}(\widehat{\bm\theta}-\bm\theta_o)$ will behave as a normal random variable as $N$ becomes large, such that the standard deviation of $\widehat{\bm\theta}$ decreases with increasing sample sizes. The Fisher information matrix can be approximated by calculating the Hessian matrix numerically.  In this way we can compute estimates of the uncertainties associated with our parameter estimates, as shown later in Section~\ref{S:Simulations}.
\subsection{Model misspecification and sampling artefacts}\label{SS:Windows}
The estimation procedures outlined in Section~\ref{SS:Whittle} can be modified to account for model misspecification as well as nonstationarity. This involves the use of semi-parametric techniques in {\em both} time and frequency, unifying various ideas in the literature.   In this section we discuss a frequency-domain semi-parametric approach for handling model misspecification and various sampling artefacts, and then address nonstationarity in the next section.

Model misspecification will occur, for example, if the variability is {\em anisotropic}, i.e. if the variance of the flow is not statistically identical in the northward and eastward directions.  Misspecification occurs because such anisotropy creates correlations between positive and negative frequencies which are ignored in equation~\eqref{e:whittlefunc2}.  To account for such possible correlations in a simple and parsimonious manner, we may choose to include only frequencies from one side of the spectrum in $\Omega$ in equation~\eqref{e:whittlefunc2}---specifically the side the inertial oscillation is known to occur.  We are then able to estimate all model parameters, including the inertial oscillation parameters as well as the background parameters appropriate for one side of the spectrum. In this way we account for the impacts of anisotropy without the need to introduce a more complicated model. 

Another reason to perform semi-parametric estimation and omit certain frequencies from the likelihood is because of various sampling artefacts. For example, surface drifter data is originally irregularly sampled in time, and is then interpolated onto a regular grid during preprocessing, see \cite{lumpkin07}. While there are many ways such an interpolation can be performed, the spectra of the resulting interpolated data are expected to differ mostly at high frequencies.  These high frequencies may also be biased by other sampling effects such as intrinsic measurement error. To account for such effects, we perform semi-parametric estimation in the frequency domain, excluding  frequencies higher than a specified cutoff. The set of modelled frequencies $\Omega$ in equation~\eqref{e:whittlefunc2} needs to include the frequency of the inertial oscillation and sufficiently many frequencies higher than this in order to resolve the damping parameter of the inertial oscillation, as well as the slope parameter of the \matern process. 
\subsection{Sample application}\label{SS:Sample}
\begin{figure}[h]
\includegraphics[width=0.49\textwidth]{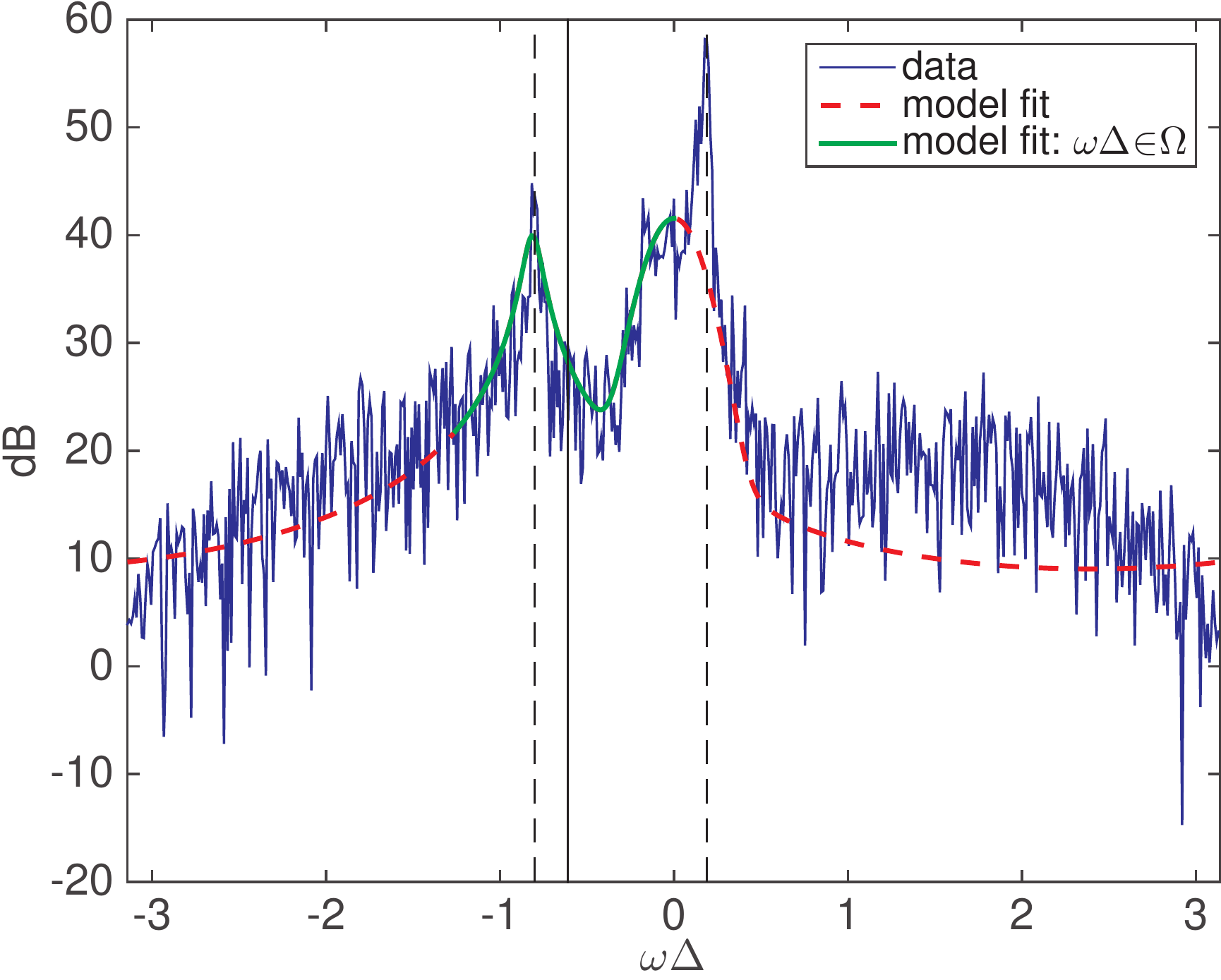}
\includegraphics[width=0.49\textwidth]{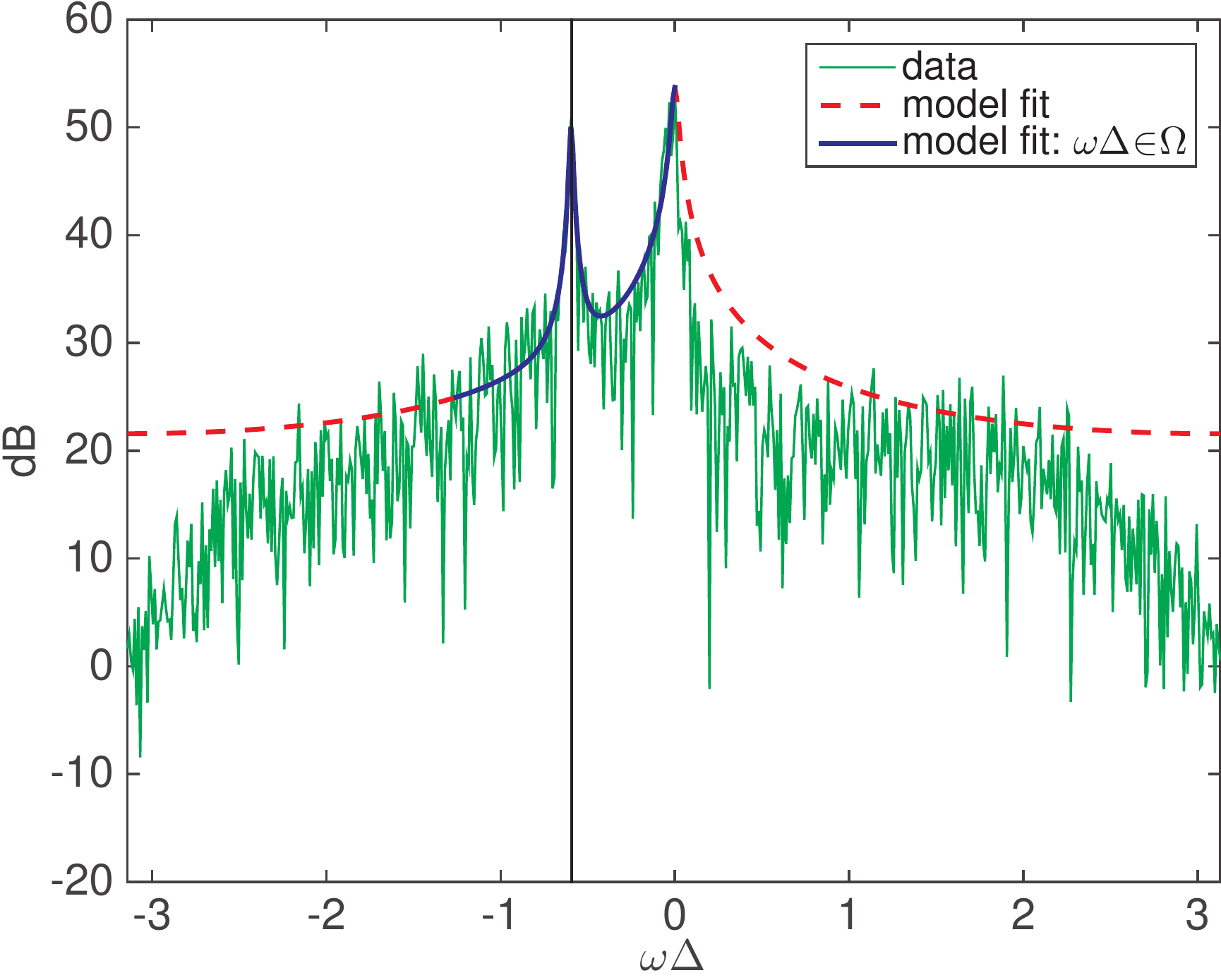}
\caption{\label{Bettyfits}Semi-parametric model fits to the blue (left) and green (right) portion of the velocity time series spectra of the drifter displayed in Figure~\ref{FigIntro}, where the vertical black solid lines denote the average Coriolis frequencies within each window.  The red-dashed lines represent the stochastic model fit extended to all frequencies.  The vertical dashed black lines on the left panel indicate the approximate location of the eddy peak at low positive frequencies $\omega_e$, and the expected shifted inertial frequency $f=f_o-\omega_e$.}
\end{figure}

As an example of the model in practice, in Figure~\ref{Bettyfits} we display semi-parametric fits, using the blurred Whittle likelihood, to the two time periods from the North Atlantic trajectory displayed in Figure~\ref{FigIntro}. This is performed with a frequency cut-off equal to $1.75f_o$, which we find to be sufficient to account for model misspecification at high frequencies, seen here as a decay associated with smoothing resulting from the interpolation process. Also, we choose to fit the model only for negative frequencies. All computations in this paper can be reproduced exactly with Matlab\textsuperscript{\textregistered} code available at www.ucl.ac.uk/statistics/research/spg/software.

During both the first and second time periods, the resulting fit with the six-parameter version of the model appears to capture the spectral structure quite well. In the second time period, the inertial peak occurs at its expected value, the local Coriolis frequency, marked by the vertical black line.  In the first time period, however, the inertial peak appears to have been shifted substantially to more negative frequencies.  The shifted amount from the local Coriolis frequency matches the frequency of a second peak occurring at low positive frequencies.  This is in agreement with the expected shift $f=f_o-\omega_e$ due to the presence of an eddy which appears to account for the low-frequency peak.

In both the left and right panels of Figure~\ref{Bettyfits}, we see that the six-parameter model appears to accurately capture the spectral structure over the modelled band of frequencies.  This is true both for a case with and without strong frequency-shifting of the inertial peak due to eddy variability, and indicates that the proposed model is indeed likely to be useful for the analysis of real-world surface drifter trajectories.  Note that in this example, a substantial asymmetry is seen to exist in the background spectra between negative and positive frequencies.  Resolution of this asymmetry is outside the scope of our model, and would require a modification.  However, the goal here is mainly to access the properties of inertial oscillations in the presence of a background flow, and this example shows that the semi-parametric approach of excluding half of the frequency axis allows us to capture the physical effects of interest within the context of a relatively simple model.

\subsection{Time-varying parameters for nonstationary time series}\label{SS:Timevarying}
Our stochastic model can also naturally be extended to accommodate nonstationarity, as will now be shown.  As drifters moves in space, they visit different regions with different spatial characteristics.  A simple example of this is that the Coriolis frequency $f_o$ is dependent on the latitude of the drifter (see equation~\eqref{e:CorrFreq}), and  thus the oscillation frequency may change as the drifter is swept across a range of latitudes.  If such effects are important, it no longer becomes natural to model the inertial oscillation using equation~\eqref{e:complexou}.  Instead we would write
\begin{equation}
\label{eqn:tvsde}
dz(t)=(i\omega_o(t)-c(t))z(t)dt+A(t)dQ(t),
\end{equation}
with time-varying coefficients. There are strong solutions to this SDE, as long as the functions $(i\omega_o(t)-c(t))z(t)$ and $A(t)$ are Borel measurable \cite[][p.~284]{karatzas1991brownian}. We may then replace the parameters of the spectrum given in equation~\eqref{e:ouspectrum} with time-varying parameters:
\begin{equation}
S^{(o)}(\omega,t)=\frac{A(t)^2}{(\omega-\omega_o(t))^2+c(t)^2}.
\label{e:outvspec}
\end{equation}
Note that using the SDE given by equation~\eqref{eqn:tvsde} does not {\em exactly} generate processes with the 
spectrum of equation~\eqref{e:outvspec}.  However, as long as $\omega_o(t)$, $c(t)$, and $A(t)$ are sufficiently smoothly varying, this difference will be a negligible effect.  See e.g. \cite{melard1989contributions} and \cite{van2006semiparametric} for more details on the mathematical theory of such time-varying processes. In any case, the  evolutionary spectrum models are useful in their own right (see for example \cite{priestley1965evolutionary}), even if they do not correspond exactly to the SDE model with time-varying coefficients.

Similarly, for the \matern process, we replace equation~\eqref{e:maternspectrum} by
\begin{equation}
\label{r-tv}
S^{(m)}(\omega,t)=\frac{B^2(t)}{(\omega^2+h^2(t))^{\alpha(t)}}.
\end{equation}
Note that a similar model was discussed by \cite{roueff2011locally} for time-varying fractional Brownian motion.  

With these time-varying models for the \matern and complex OU processes, we can then estimate the time-varying parameters by considering a rolling window of observations from the overall time series.  Specifically, we define a window length $W$, with the model parameters $\widehat{\bm\theta}(t)$ being estimated as a function of time using the observations $\{Z_{t-W/2+1},\ldots,Z_t,$ $\ldots,Z_{t+W/2}\}$ from equations~\eqref{eq:periodogram} and \eqref{e:whittlefunc2}.  This procedure to capture time-varying parameters is semi-parametric, in that the locally stationary process has a parametric model, but no model is imposed on the evolution of each parameter value over time globally.  This approach is similar in spirit to \cite{dahlhaus1997fitting} and can be understood as a special case of local inference using an empirical spectral measure, with an appropriate choice of kernel function, see also \cite{dahlhaus2009local}.

In order for this time-varying estimation to yield accurate estimates of the model parameters, it must be possible to choose the window length $W$ so that the model parameters vary little over the window timescale, with $W$ still large enough so that there is sufficient averaging to reduce variance of the parameter estimates.   For example, numerical explorations have shown that a drift in the inertial oscillation frequency over the window will result in an overestimate of the damping parameter~$c$.  This happens as the estimated energy of the inertial oscillation is spread around the true value at time $t$ and gives the appearance of a flatter, broader peak around the average inertial frequency.  We examine such issues empirically through application to data in Section~\ref{S:Simulations}.  A more formal investigation of how time variation of the parameters impacts the estimation errors is outside the scope of this study.
\subsection{Model choice and likelihood ratio tests}\label{SS:Modelchoice}
In Section~\ref{SS:Complete}, we discussed that it may be valuable to compare the full, six-parameter stochastic model against various reduced versions.  We may select between these different variants of our stochastic model, and their corresponding parameter estimates,  using likelihood ratio tests.  The methodology of  likelihood ratio tests for complex-valued time series is detailed in \cite{sykulski2013whittle}.  In summary, to perform the test at time $t$, we use the likelihood ratio test statistic given by
\begin{equation}
R(t)=2(\ell_f(\widehat{\bm\theta}_1{(t)})-\ell_f(\widehat{\bm\theta}_o{(t)}))
\label{e:ratiotest}
\end{equation}
where $\widehat{\bm\theta}_o{(t)}$ is the fitted parameter under the null hypothesis and $\widehat{\bm\theta}_1{(t)}$ is the fitted parameter under the alternative model.  The likelihood ratio statistic $R(t)$ is found to be asymptotically distributed according to a $\chi^2$ distribution with degrees of freedom equal to the number of extra parameters in the alternative hypothesis versus the null.  Note that if we wish to compare more than two nested models at a time from Section~\ref{SS:Complete} then we can use model choice procedures appropriate for complex-valued time series, as developed in \cite{sykulski2013whittle}.
\section{Application to Data}\label{S:Simulations}
In this section we test and apply our stochastic model to an ensemble of drifter trajectories from numerical model output in Section~\ref{SS:Numerical}, and to real-world data from the Global Drifter Program in Section~\ref{SS:Real}.  The results indicate that our model is able to accurately capture the observed structure, supporting the use of this analysis method as a promising tool for physical inquiry. All data and statistical computations performed can be reproduced exactly with Matlab\textsuperscript{\textregistered} code available at www.ucl.ac.uk/statistics/research/spg/software. Included in this online material is a simple routine for estimating parameters of time series as a \matern and/or complex-OU process, which can be adjusted accordingly for other stochastic processes.

\subsection{Numerical Data}\label{SS:Numerical}
We start by investigating the performance of our proposed inference method on output from an idealised numerical model of ocean circulation. Synthetic surface drifter trajectories were generated using a wind-forced S-coordinate Primitive Equation Model (SPEM 5.2), with settings very similar to those of \cite{danioux2008propagation}. The spatially homogenous surface wind forcing was based on an observed wind time series that has previously been used as the forcing field in the numerical study of \cite{klein2004wind}.  The simulation is run without wind until a near steady-state in the background turbulence is achieved, after which the wind forcing is turned on. At this point the model is seeded with 200 near-surface particles, spread uniformly in the active region of the domain. The trajectories are then subsampled at two hour resolution resulting in 200 trajectories each with length 1,200 time points. A more detailed description of the numerical model can be found within the online code.

\begin{figure}[h]
\includegraphics[width=0.49\textwidth]{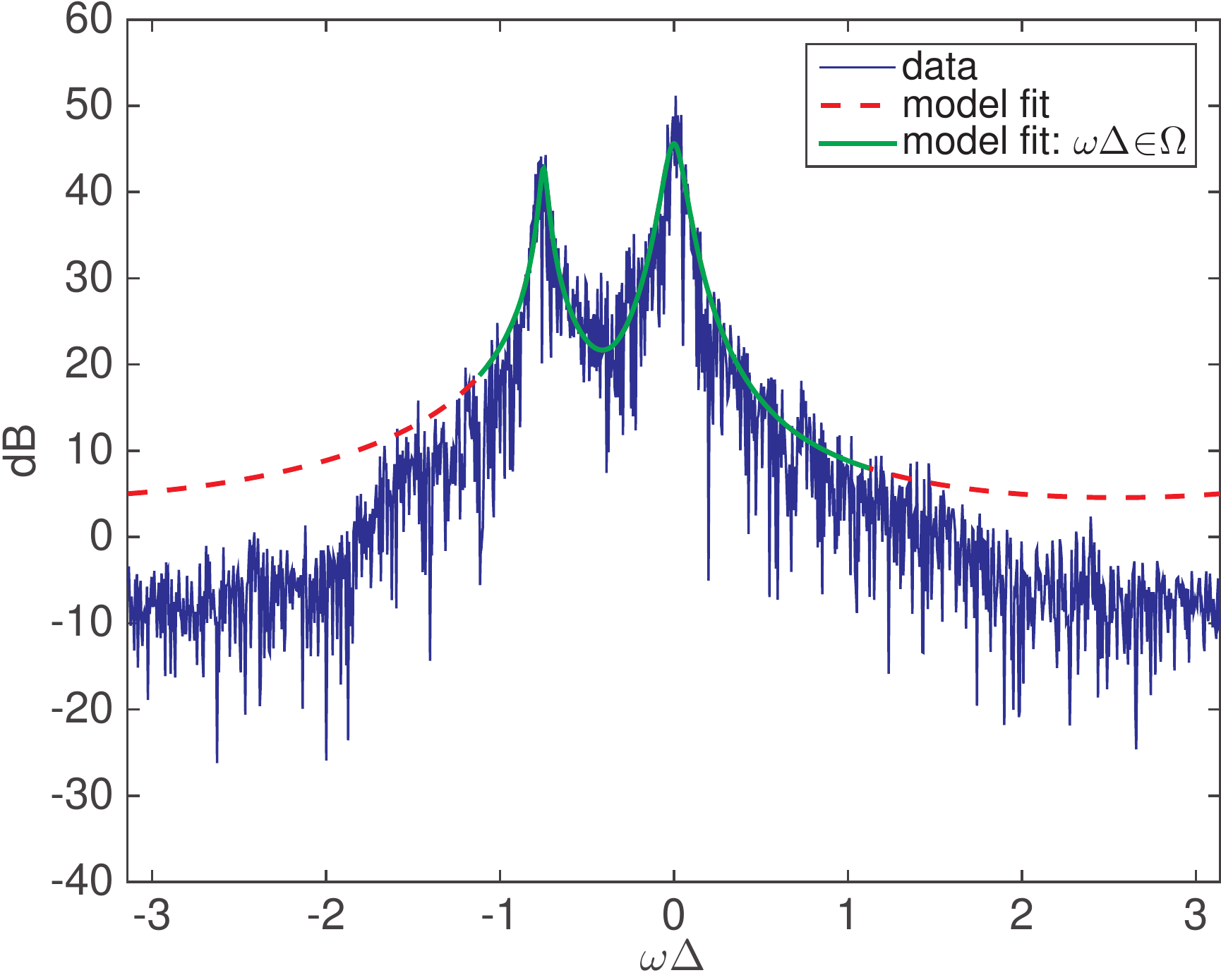}
\includegraphics[width=0.49\textwidth]{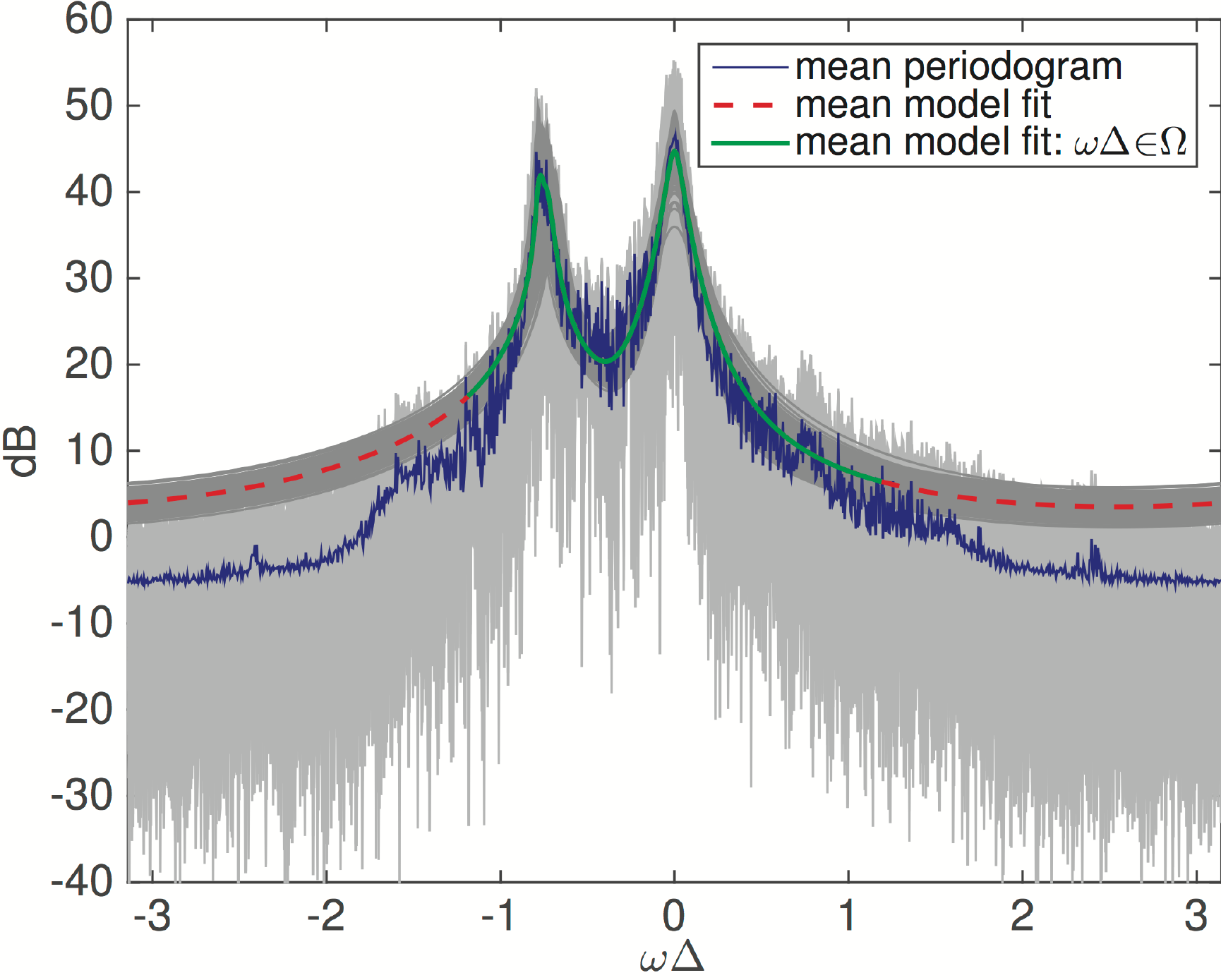}
\caption{\label{EricFigs}An individual (left) and the ensemble mean (right) periodograms from the numeral model output, together with our parametric fits.  The individual and mean model fit are overlaid in green for the frequencies used in the semi-parametric estimation, and extended outwards in red for all other observed frequencies.  On the right figure in greyscale are the entire ensemble of observed (light grey) and modelled (dark grey) spectra.  Spectral energy is represented on a decibel scale.}
\end{figure}

We fit our 6-parameter stochastic model to each time series.  For this data we fit the stochastic model to {\em both} sides of the spectra, as the numerical model is largely devoid of strong eddies and other motions which could cause substantial asymmetry in the rotary spectrum, apart from the inertial oscillations themselves.  We do however perform a semi-parametric fit in the frequency domain, by omitting frequencies higher than $1.5f_o$ in the optimisation.  Because the time series are relatively short, we assume the parameters are not time-varying for each time series.  An individual fit is displayed in Figure~\ref{EricFigs} (left).  Then in Figure~\ref{EricFigs} (right) we display the mean periodogram and mean fit to all 200 time series, with individual fits and spectral estimates shown as grey shading.  Note that in both plots the vertical axis is in decibels to show multiscale effects, while the horizontal axis is linear.  

The stochastic model captures the variability of the numerical model output consistently well, particularly the inertial oscillation, which does not exhibit substantial shifting in this data.  Furthermore, the spread of model fits is narrow (dark grey shading), indicating that the model can naturally account for the much larger variability in the data periodograms (light grey shading) with only a small range of parameter variations.  Some variability in the model fits is expected to remain as the drifters are seen to experience different local spatial conditions within the numerical simulation.  While there does appear to be some slight asymmetry in the background spectrum that is not captured by our model, this is not apparent until relatively high frequencies where the spectral levels have decayed substantially from their peak values. 

\begin{figure}
\includegraphics[width=0.49\textwidth]{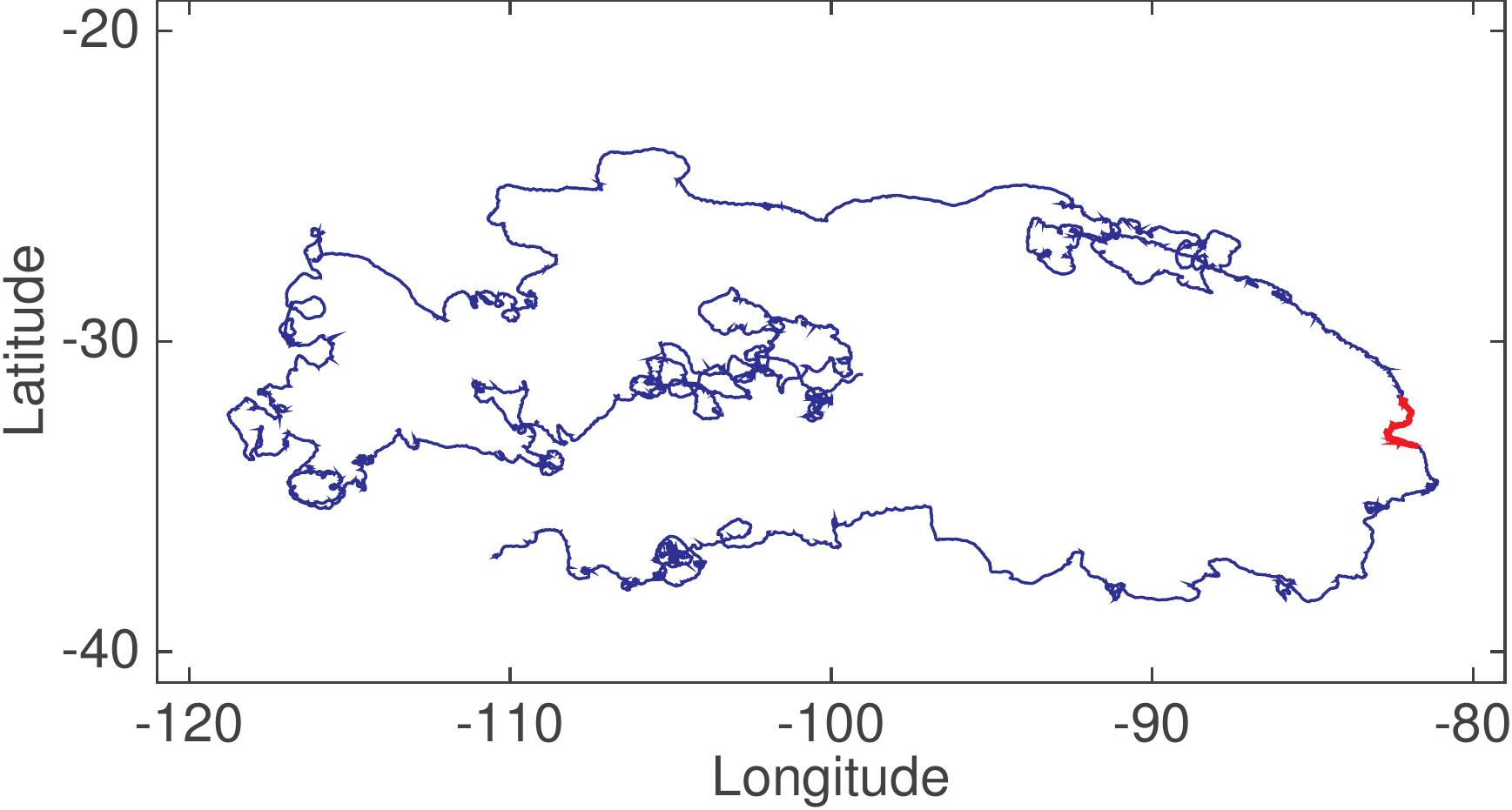}
\includegraphics[width=0.49\textwidth]{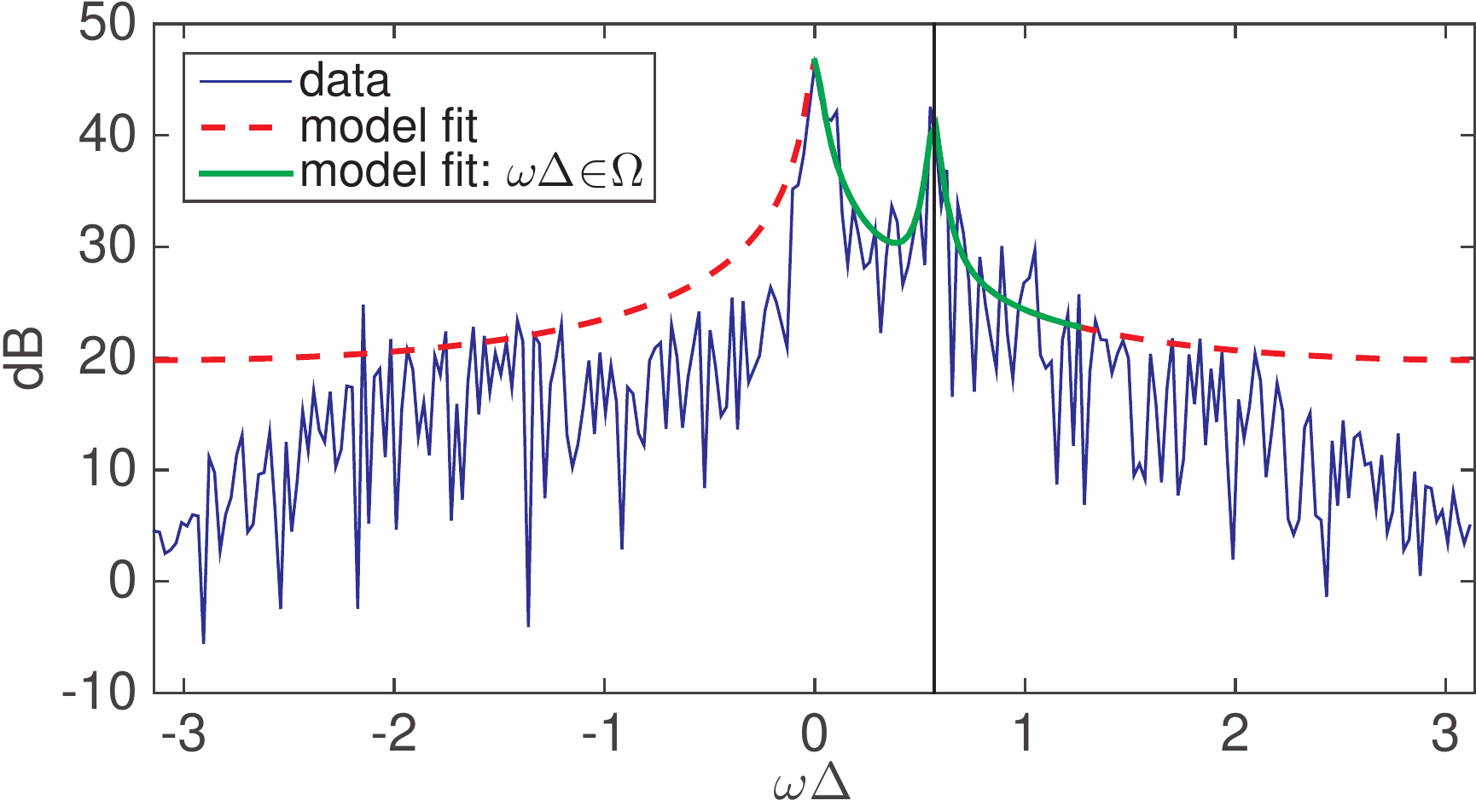}
\caption{\label{RealPlots}A 1,642-day trajectory of a South Pacific drifter (left), the portion highlighted in red is from day 358 to 378. The spectral estimate and stochastic model fit to this 20-day velocity time series is shown (right), with the average Coriolis frequency in this window shown as a black vertical line.}
\vspace{-3mm}
\end{figure}

\subsection{Real Drifter Data}\label{SS:Real}
In this section we apply our stochastic model to a 1,642-day trajectory of a South Pacific drifter, Global Drifter Program ID~\#44312, shown in Figure~\ref{RealPlots} (left).  A one-sided model fit to a 244-point portion of this data (corresponding to a 20-day trajectory from day 358 to 378 from the start of the series) is displayed in Figure~\ref{RealPlots} (right).  The inertial oscillation and turbulent background have been captured accurately in the modelled frequencies, but there is again an asymmetry between negative and positive frequencies, motivating our semi-parametric approach in frequency.  Note that inertial oscillations occur at positive frequencies for this Southern Hemisphere drifter.

\begin{figure}[h]
\centering
\includegraphics[width=0.913\textwidth]{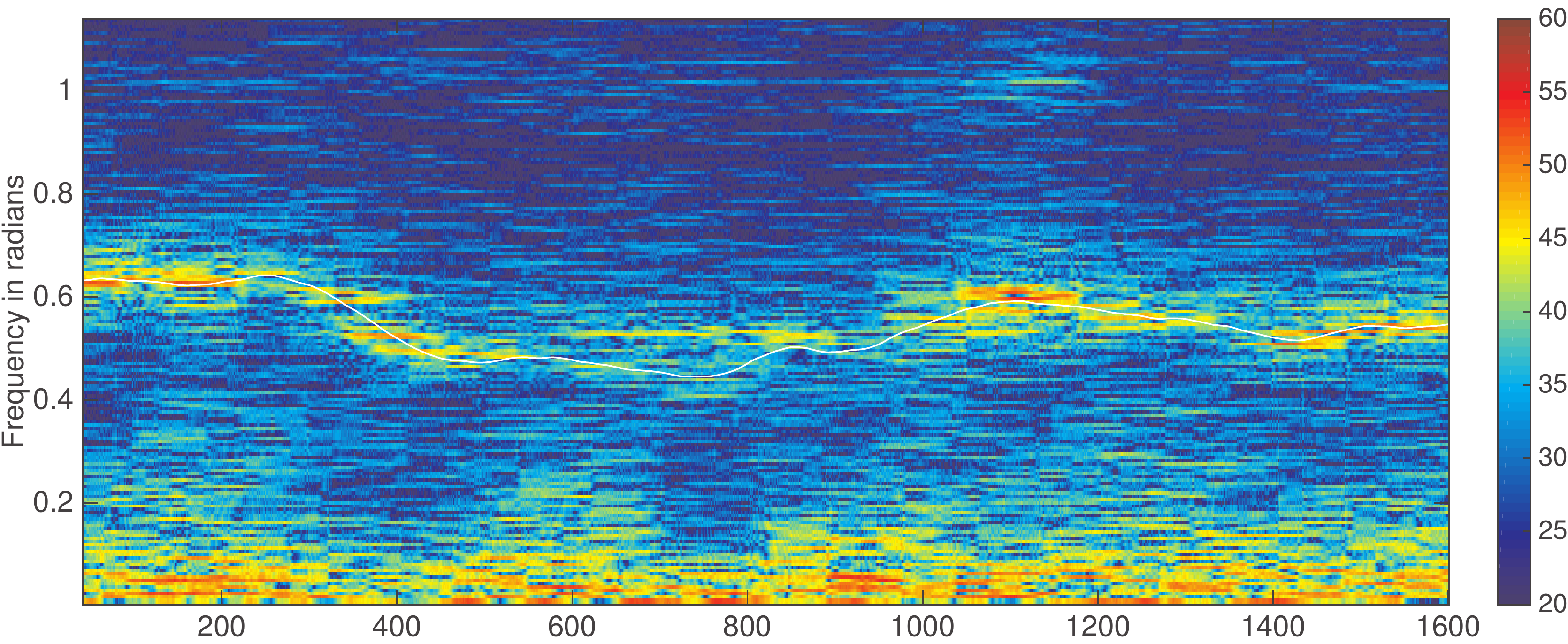}
\includegraphics[width=0.913\textwidth]{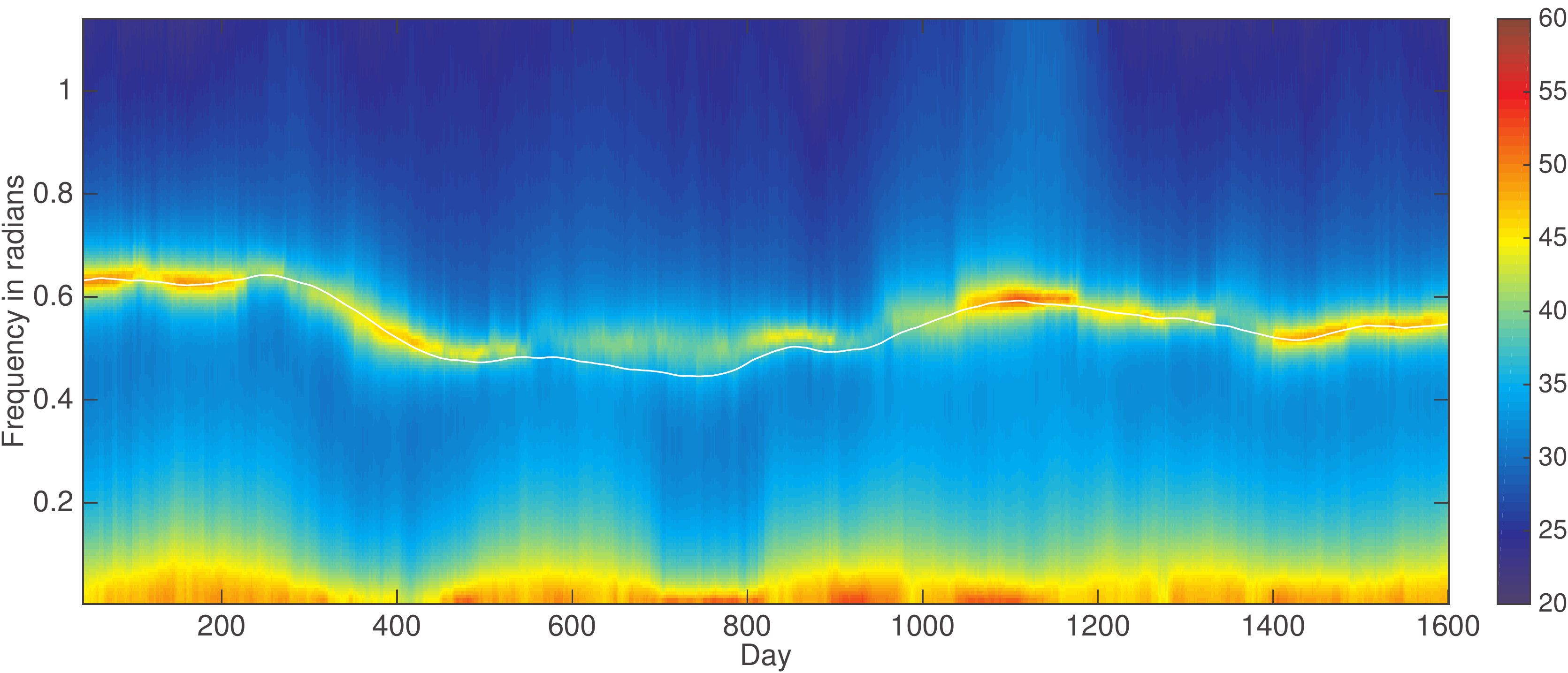}
\caption{\label{ImageScale2}Time-varying spectra of the South Pacific drifter displayed in Figure~\ref{RealPlots} (on a decibel scale, where red is high energy, blue is low energy), where the spectra of observed data is displayed above, and the modelled spectra using our stochastic model is below. The local Coriolis frequency is represented with a white line. We only display the range of frequencies considered in the estimation of parameters.}
\end{figure}

This time series will be used to illustrate our time-varying parametric estimation procedure discussed in Section~\ref{SS:Timevarying}. In Figure~\ref{ImageScale2} we display image-scale plots of the observed (top) and modelled (bottom) log-spectra at each time point, resulting form a rolling-window fit with a window length $W=1,000$ time points (or 83.3 days), focusing on the frequencies included in the semi-parametric likelihood estimation.   During most of the time series, the model identifies the inertial frequency as occurring at the local Coriolis frequency (shown as the white line); note that smooth variation of the estimated inertial frequency is observed,  even though our optimisation procedure is unconstrained for this parameter.  In general, our model captures the temporal variability observed in the moving-window periodogram at both the inertial peak and the low-frequency peak, and this good agreement shows that structure in the data can be well accounted for by our simple model.

\begin{figure}
\centering
\includegraphics[width=0.913\textwidth]{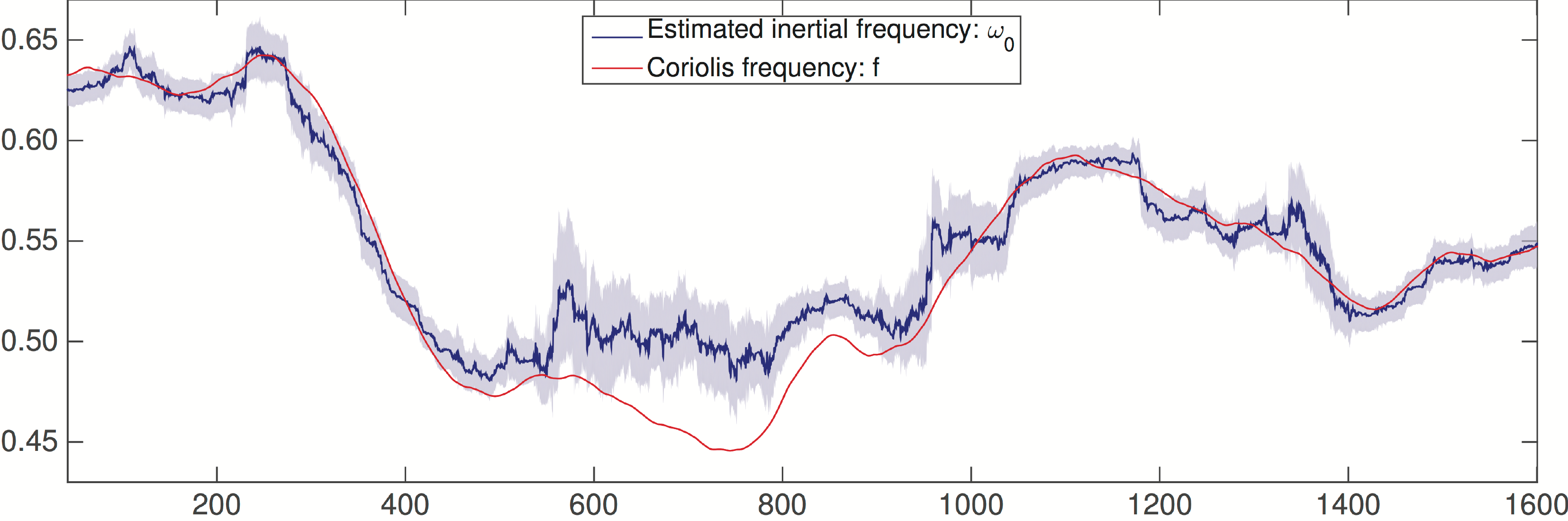}\vspace{2mm}
\includegraphics[width=0.913\textwidth]{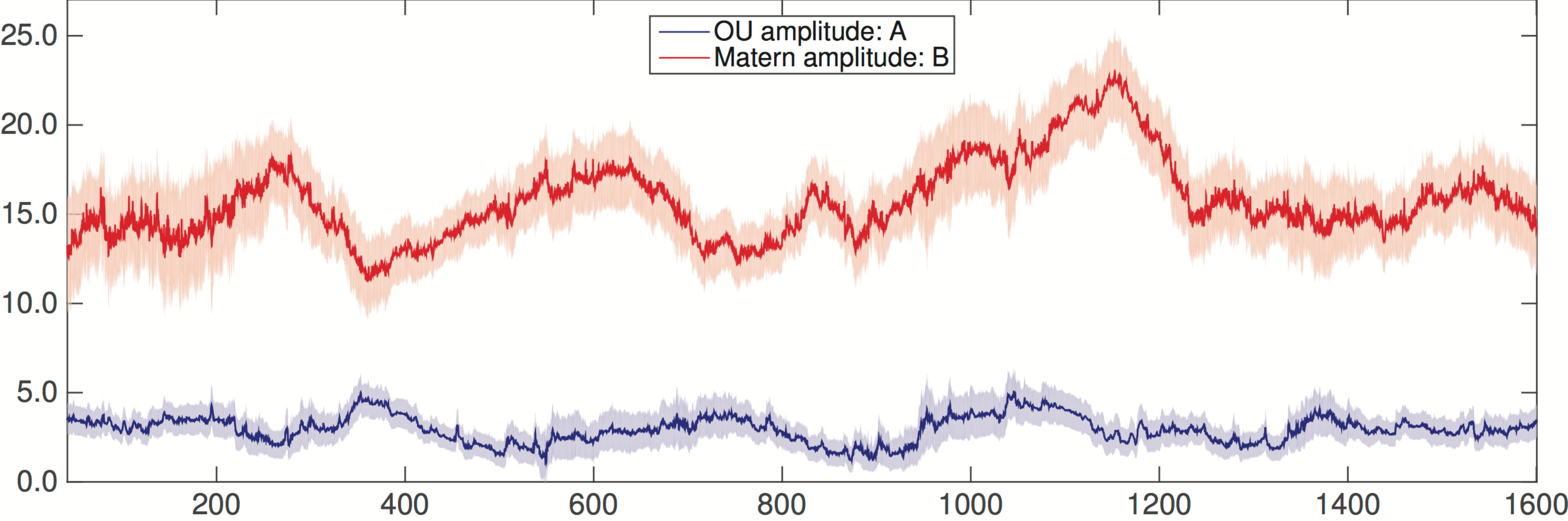}\vspace{2mm}
\includegraphics[width=0.913\textwidth]{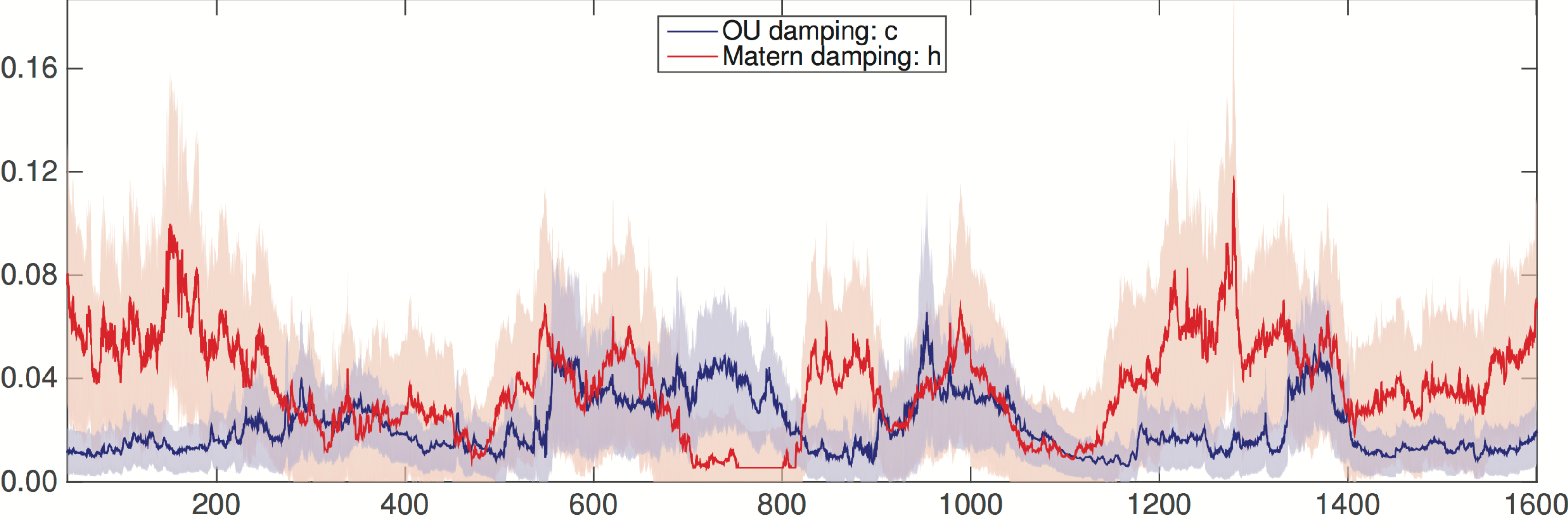}\vspace{2mm}
\includegraphics[width=0.913\textwidth]{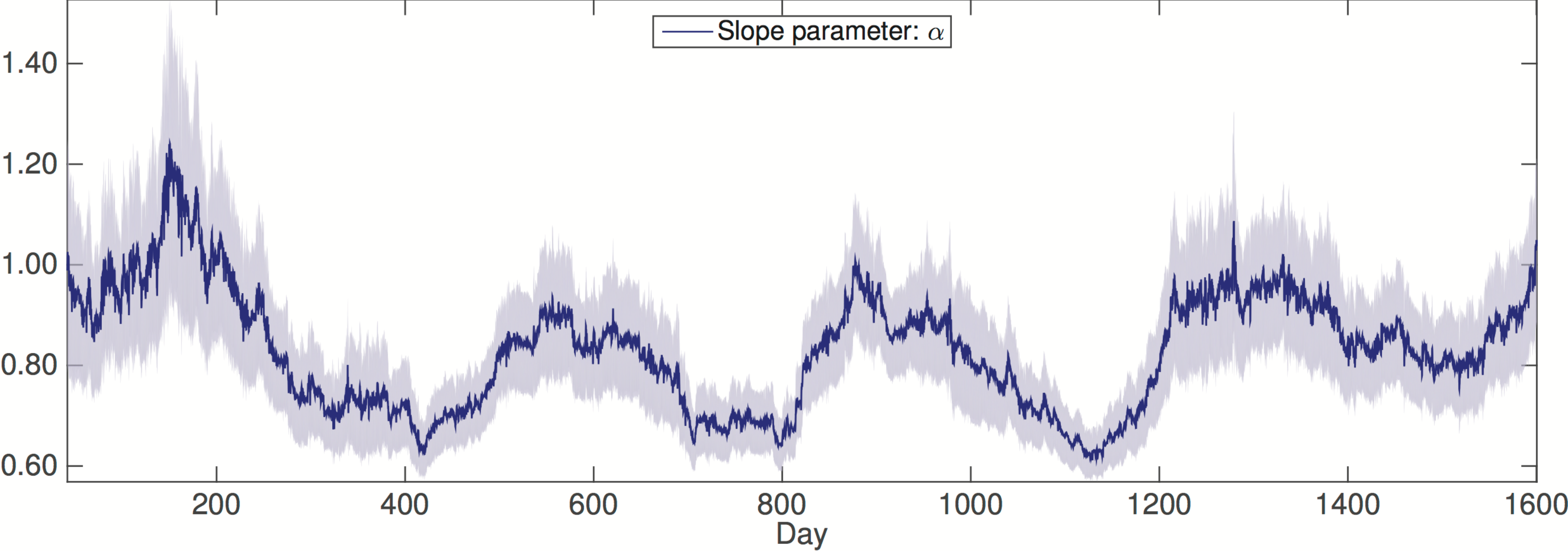}
\caption{\label{Parameters}Time-varying parameter values of the model fits of Figure~\ref{ImageScale2}, with 95\% confidence intervals calculated from the Hessian matrix.}
\end{figure}

In Figure~\ref{Parameters} we display the estimates of the six time-varying parameters in our model, which evolve over time relatively smoothly, even though our semi-parametric model does not specify how these parameters should change over time.  We also include 95\% confidence intervals which are calculated using equation~\eqref{eq:fisher}, where we approximate the Fisher information matrix using the observed Hessian matrix \citep{efron1978assessing}. Here we see more clearly that the inertial oscillation is shown to significantly shift from the theoretical frequency particularly between days $600$ and $900$.  Inspecting Figure~\ref{ImageScale2} in more detail, we can see that there are two bands of energy during this time period near the Coriolis frequency.

A closer examination (not shown) reveals that the higher frequency band is in fact attributed to tidal energy, and occurs during a time period when the drifter passes over a relatively shallow region of the ocean.  Our stochastic model has selected this frequency for the inertial peak, and has used a higher damping parameter $c$ (see Figure~\ref{Parameters}) to then capture effects from both energy bands.  Another artefact from tides can be seen around day $1,150$ where there is semidiurnal tidal energy near frequency $\omega\Delta=1$, see Figure~\ref{ImageScale2}.  Our model has captured this energy with the \matern process using an unusually low slope parameter $\alpha$ (see Figure~\ref{Parameters}).  While our model is apparently misspecified in these regions, this is because of tidal effects which have not been correctly removed from the data. In addition to correctly identifying frequency shifts from the local Coriolis frequency due to eddies, as well as the expected variation of the Coriolis frequency with latitude, our methods can therefore also flag portions of drifter trajectories which require separate investigation or additional analysis.

\begin{figure}[h]
\centering
\includegraphics[width=0.913\textwidth]{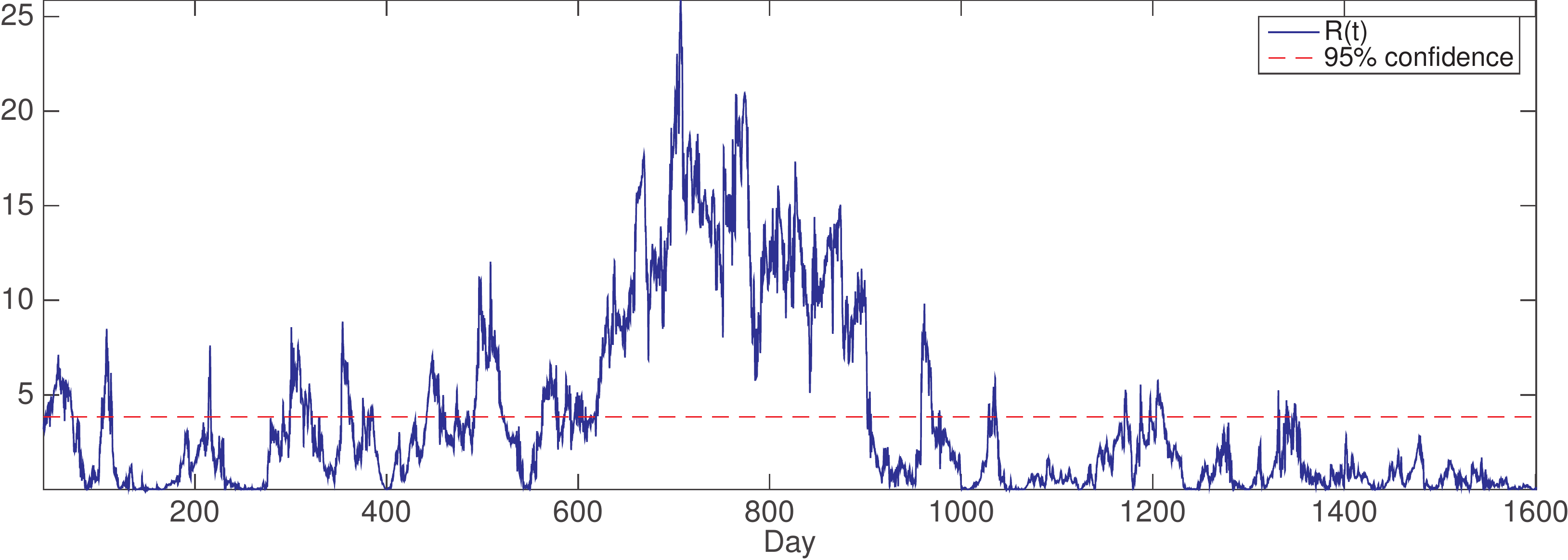}
\caption{\label{RatioTest}The likelihood ratio test statistic over time, comparing the full 6-parameter model with the null 5-parameter model where $\omega_o=f$, the Coriolis frequency.}
\end{figure}

As discussed in Section~\ref{SS:Modelchoice}, we can also use likelihood ratio tests to select between variants of our model described in Section~\ref{SS:Complete}.  To this end, we investigate the possibility of a shifted inertial frequency from the local Coriolis frequency $f_o$.  We therefore also fit a simpler 5-parameter model to this data where we fix the inertial frequency $\omega_o=f_o$.  The likelihood ratio test statistic from the two models at each time point is displayed in Figure~\ref{RatioTest}.  Values above the 95\% significance level indicate that the null 5-parameter model should be rejected in favour of the full 6-parameter model with shifted frequency $\omega_o\neq f_o$.  The results are consistent with the confidence intervals found (via the Hessian matrix) in Figure~\ref{Parameters}, with a significant shift occurring between days $600$ and $900$ due to tidal effects.

In our analysis we note that we have used a relatively long window size of 1,000 time points, corresponding to 83.3 days.  This is because we have deliberately reduced variance, at the expense of bias,  for this single-drifter analysis.  In general, the use of shorter windows will be favoured, such as in the right panel of Figure~\ref{RealPlots}, when multiple trajectories are available within a given spatial region.  Bias can then be reduced, and the variance is instead reduced by averaging over the multiple time series.  This yields parameter output with high spatial resolution, which is important to oceanographers in resolving spatial heterogeneity at small and large scales in the oceans.

\begin{figure}[h]
\centering
\includegraphics[width=0.8\textwidth]{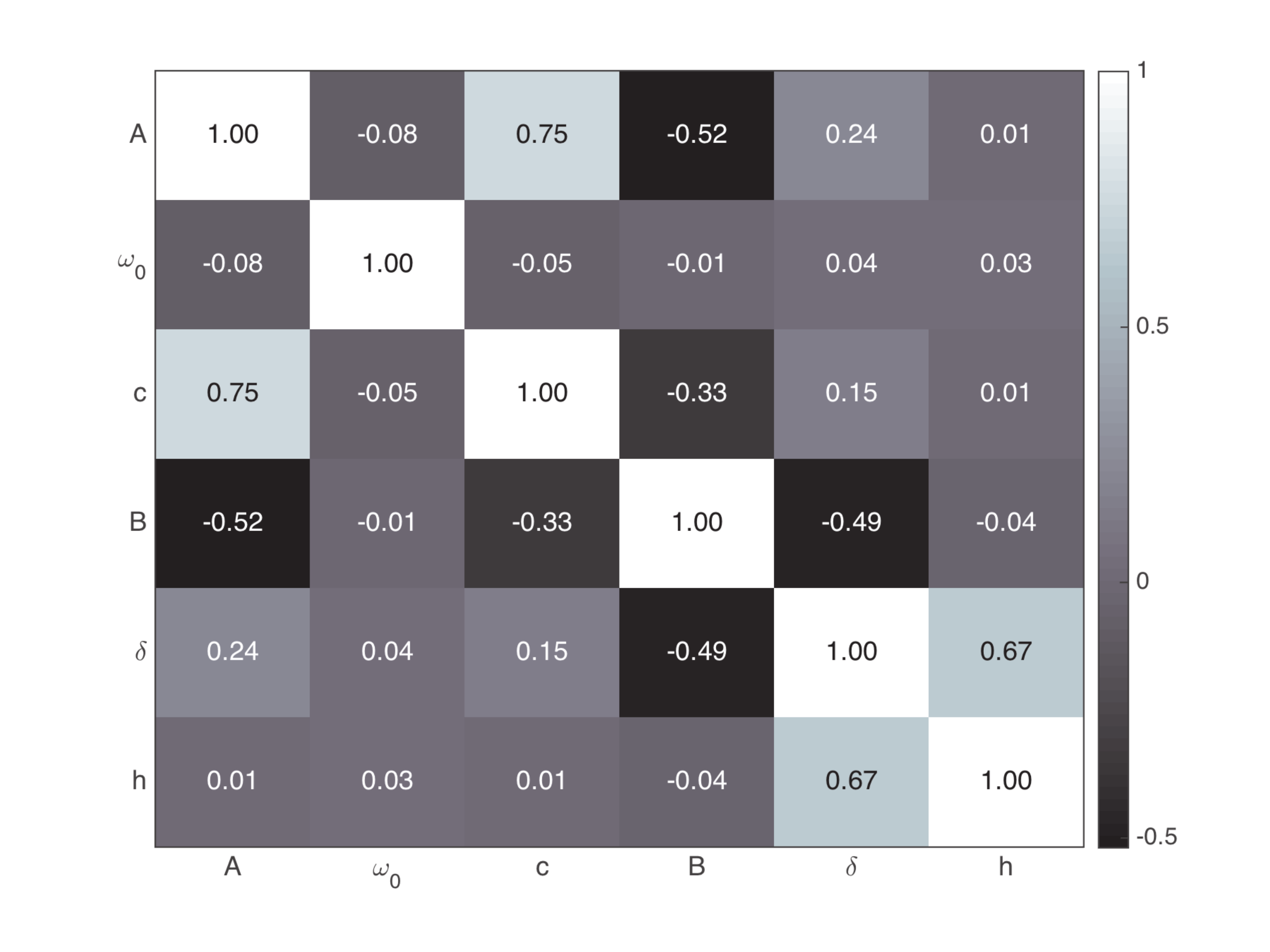}
\vspace{-3mm}
\caption{\label{Hessian2}Average median correlation of each pair of parameter estimates, estimated using equation~\eqref{eq:fisher} and the observed Hessian matrix.}
\end{figure}

The Hessian matrix also allows us to compute the correlation between the estimates of each parameter over time.  Figure~\ref{Hessian2} displays the average correlation between each pair of parameters.  There is strong correlation (white for positive and black for negative) within the background parameters ($\widehat{B}(t)$, $\widehat{\alpha}(t)$ and $\widehat{h}(t)$) and between the damping and amplitude of the inertial oscillation ($\widehat{c}(t)$ and $\widehat{A}(t)$) as expected.  Note also that the estimates of the inertial frequency, $\widehat{\omega}_o(t)$, are largely uncorrelated with the other parameter estimates.  This is also expected as $\widehat{\omega}_o(t)$ is the only parameter that does not affect either the shape or the amplitude of the spectral model, only the frequency location of the inertial peak. In addition, as we would expect to estimate $\omega_o(t)$ with better precision than the other parameters (see \cite{hannan1986law} for example), we would therefore expect their codependence to decay quickly. Finally, we note that there is some negative correlation between the amplitude of the background, $\widehat{B}(t)$, and the inertial parameters $\widehat{A}(t)$ and $\widehat{c}(t)$, which suggests that the background and inertial oscillation cannot be modelled separately and need to be combined in one estimation procedure, as performed in this paper.  For example, it would be na\"{\i}ve to only use frequencies near the Coriolis force to estimate inertial parameters and ignore the background---as there is still considerable background energy at these frequencies, as suggested by the correlations between the parameter estimates.
\vspace{-3mm}
\section{Conclusion}\label{S:Conclusion}
In this paper we have constructed a statistical model for analysing and interpreting ocean surface drifter trajectories.  This is the first stochastic model to accommodate the effects of inertial oscillations together with the turbulent background process.  The inertial oscillations are modelled as a complex Ornstein-Uhlenbeck (OU) process, motivated by our physical understanding of their structure.  The turbulent background is modelled flexibly using a \matern process---which encompasses and generalises existing models in the literature, and which we have shown to exhibit the advantages of fractional Brownian motion without suffering from the serious drawback of unbounded energy with increasing time.

Our stochastic model is simple, with only six parameters, and we have demonstrated how these parameters can be estimated from data with spatial and temporal variability by allowing the parameters to be time-varying and estimated semi-parametrically over rolling windows. We also showed how to correct for sampling issues such as aliasing and irregular temporal sampling.  We tested our modelling and estimation techniques on real-world and numerically generated trajectories with favourable results, showing that the model is accurate, robust to misspecifications, and able to track changes over time and space. By modelling processes carefully in terms of components, we are able to reconstruct the variety of effects encountered in the data, and also explain a number of the observed artefacts or other features that are not captured by our model. 

The model development and testing carried out herein is a necessary step before carrying out an analysis of the full dataset from the Global Drifter Program.   Summary maps created from this application would be expected to yield new insights, and potentially impact the value of the oceanographic surface drifter data to the climate modelling effort \citep[e.g.][]{mcclean08-omer}.  Another direction of future work is to explicitly incorporate models for detecting and extracting eddies \citep{lilly11-grl}, so that our stochastic model can also be applied to drifters contaminated with eddies.  In agreement with the theory of \cite{kunze85-jpo}, an example of strong shifting of an inertial oscillation by an eddy was found.  This particularly interesting feature deserves further attention, and also raises the question as to how many other similar frequency shifts are to be found in the global drifter dataset, an issue that may be significant for the surface energy balance of the oceans \citep{elipot10-jgr}.


\begin{thebibliography}{55}
\expandafter\ifx\csname natexlab\endcsname\relax\def\natexlab#1{#1}\fi
\expandafter\ifx\csname url\endcsname\relax
  \def\url#1{\texttt{#1}}\fi
\expandafter\ifx\csname urlprefix\endcsname\relax\def\urlprefix{URL}\fi

\bibitem[{Abramowitz and Stegun(1972)}]{abramowitz1964handbook}
Abramowitz, M. and Stegun, I.~A. (1972) \textit{Handbook of mathematical
  functions with formulas, graphs, and mathematical tables}.
\newblock National Bureau of Standards.

\bibitem[{Arat\'o et~al.(1962)Arat\'o, Kolmogorov and
  Sinai}]{Arato1962estimation}
Arat\'o, M., Kolmogorov, A.~N. and Sinai, Y.~G. (1962) Estimation of the
  parameters of a complex stationary {G}aussian {M}arkov process.
\newblock \textit{Dokl. Akad. Nauk SSSR}, \textbf{146}, 747--750.

\bibitem[{Beran(1994)}]{beran1994statistics}
Beran, J. (1994) \textit{Statistics for long-memory processes}, vol.~61.
\newblock New York: Chapman and Hall.

\bibitem[{Berloff and McWilliams(2002)}]{berloff2002material}
Berloff, P.~S. and McWilliams, J.~C. (2002) Material transport in oceanic
  gyres. {P}art {II}: Hierarchy of stochastic models.
\newblock \textit{J. Phys. Oceanogr.}, \textbf{32}, 797--830.

\bibitem[{Bowman et~al.(2009)Bowman, Giannitrapani and
  Marian}]{bowman2009spatiotemporal}
Bowman, A.~W., Giannitrapani, M. and Marian, S.~E. (2009) Spatiotemporal
  smoothing and sulphur dioxide trends over {E}urope.
\newblock \textit{J. R. Statist. Soc. C}, \textbf{58}, 737--752.

\bibitem[{Brockwell et~al.(2007)Brockwell, Davis and
  Yang}]{brockwell2007continuous}
Brockwell, P.~J., Davis, R.~A. and Yang, Y. (2007) Continuous-time {G}aussian
  autoregression.
\newblock \textit{Statist. Sin.}, \textbf{17}, 63--80.

\bibitem[{Dahlhaus(1997)}]{dahlhaus1997fitting}
Dahlhaus, R. (1997) Fitting time series models to nonstationary processes.
\newblock \textit{Ann. Statist.}, \textbf{25}, 1--37.

\bibitem[{Dahlhaus(2009)}]{dahlhaus2009local}
--- (2009) Local inference for locally stationary time series based on the
  empirical spectral measure.
\newblock \textit{J. Econometrics}, \textbf{151}, 101--112.

\bibitem[{Danioux et~al.(2008)Danioux, Klein and
  Rivi{\`e}re}]{danioux2008propagation}
Danioux, E., Klein, P. and Rivi{\`e}re, P. (2008) Propagation of wind energy
  into the deep ocean through a fully turbulent mesoscale eddy field.
\newblock \textit{J. Phys. Oceanogr.}, \textbf{38}, 2224--2241.

\bibitem[{Davis et~al.(2013)Davis, Kl{\"u}ppelberg and
  Steinkohl}]{davis2013statistical}
Davis, R.~A., Kl{\"u}ppelberg, C. and Steinkohl, C. (2013) Statistical
  inference for max-stable processes in space and time.
\newblock \textit{J. R. Statist. Soc. B}, \textbf{75}, 791--819.

\bibitem[{Dzaparidze(1974)}]{dzaparidze1974new}
Dzaparidze, K.~O. (1974) A new method for estimating spectral parameters of a
  stationary regular time series.
\newblock \textit{Theor. Probab. Appl.}, \textbf{19}, 122--132.

\bibitem[{Dzhaparidze and Kotz(1986)}]{dzhaparidze1986parameter}
Dzhaparidze, K.~O. and Kotz, S. (1986) \textit{Parameter estimation and
  hypothesis testing in spectral analysis of stationary time series}.
\newblock Springer-Verlag.

\bibitem[{Early(2012)}]{early2012forces}
Early, J.~J. (2012) The forces of inertial oscillations.
\newblock \textit{Q. J. Roy. Meteor. Soc.}, \textbf{138}, 1914--1922.

\bibitem[{Efron and Hinkley(1978)}]{efron1978assessing}
Efron, B. and Hinkley, D.~V. (1978) Assessing the accuracy of the maximum
  likelihood estimator: Observed versus expected {F}isher information.
\newblock \textit{Biometrika}, \textbf{65}, 457--483.

\bibitem[{Elipot et~al.(2010)Elipot, Lumpkin and Prieto}]{elipot10-jgr}
Elipot, S., Lumpkin, R. and Prieto, G. (2010) Modification of inertial
  oscillations by the mesoscale eddy field.
\newblock \textit{J. Geophys. Res.}, \textbf{115}, {C}09010 (1--20).

\bibitem[{Erhardt et~al.(2012)Erhardt, Allen, Wei, Eichele and
  Calhoun}]{erhardt2012simtb}
Erhardt, E.~B., Allen, E.~A., Wei, Y., Eichele, T. and Calhoun, V.~D. (2012)
  Sim{TB}, a simulation toolbox for f{MRI} data under a model of spatiotemporal
  separability.
\newblock \textit{Neuroimage}, \textbf{59}, 4160--4167.

\bibitem[{Fuentes et~al.(2013)Fuentes, Henry and
  Reich}]{fuentes2013nonparametric}
Fuentes, M., Henry, J. and Reich, B. (2013) Nonparametric spatial models for
  extremes: application to extreme temperature data.
\newblock \textit{Extremes}, \textbf{16}, 75--101.

\bibitem[{Gille(2005)}]{gille05-jaot}
Gille, S.~T. (2005) Statistical characterization of zonal and meridonal ocean
  wind stress.
\newblock \textit{J. Atmos. Ocean Tech.}, \textbf{22}, 1353--1372.

\bibitem[{Gneiting et~al.(2010)Gneiting, Kleiber and
  Schlather}]{gneiting2010matern}
Gneiting, T., Kleiber, W. and Schlather, M. (2010) Mat{\'e}rn cross-covariance
  functions for multivariate random fields.
\newblock \textit{J. Am. Statist. Ass.}, \textbf{105}, 1167--1177.

\bibitem[{Gneiting et~al.(2012)Gneiting, {\v{S}}ev{\v{c}}{\'\i}kov{\'a} and
  Percival}]{gneiting2012estimators}
Gneiting, T., {\v{S}}ev{\v{c}}{\'\i}kov{\'a}, H. and Percival, D.~B. (2012)
  Estimators of fractal dimension: assessing the roughness of time series and
  spatial data.
\newblock \textit{Statist. Sci.}, \textbf{27}, 247--277.

\bibitem[{Griffa et~al.(2007)Griffa, Kirwan, Mariano, {\"O}zg{\"o}kmen and
  Rossby}]{griffa2007lagrangian}
Griffa, A., Kirwan, A.~D., Mariano, A.~J., {\"O}zg{\"o}kmen, T. and Rossby, T.
  (2007) \textit{{L}agrangian analysis and prediction of coastal and ocean
  dynamics}.
\newblock Cambridge University Press.

\bibitem[{Guinness and Stein(2013)}]{guinness2013interpolation}
Guinness, J. and Stein, M.~L. (2013) Interpolation of nonstationary high
  frequency spatial--temporal temperature data.
\newblock \textit{Ann. Appl. Statist.}, \textbf{7}, 1684--1708.

\bibitem[{Hannan and Mackisack(1986)}]{hannan1986law}
Hannan, E.~J. and Mackisack, M. (1986) A law of the iterated logarithm for an
  estimate of frequency.
\newblock \textit{Stoch. Proc. Appl.}, \textbf{22}, 103--109.

\bibitem[{Herrera and Bayen(2010)}]{herrera2010incorporation}
Herrera, J.~C. and Bayen, A.~M. (2010) Incorporation of {L}agrangian
  measurements in freeway traffic state estimation.
\newblock \textit{Transport. Res. B-Meth.}, \textbf{44}, 460--481.

\bibitem[{Jeffreys(1942)}]{jeffreys1940variation}
Jeffreys, H. (1942) The variation of latitude.
\newblock \textit{Mon. Not. R. Astron. Soc.}, \textbf{100}, 139--155.

\bibitem[{Karatzas and Shreve(1991)}]{karatzas1991brownian}
Karatzas, I.~A. and Shreve, S.~E. (1991) \textit{Brownian motion and stochastic
  calculus}, vol. 113.
\newblock Springer.

\bibitem[{Klein et~al.(2004)Klein, Lapeyre and Large}]{klein2004wind}
Klein, P., Lapeyre, G. and Large, W.~G. (2004) Wind ringing of the ocean in
  presence of mesoscale eddies.
\newblock \textit{Geophys. Res. Lett.}, \textbf{31}, {L}15306.

\bibitem[{Klocker et~al.(2012)Klocker, Ferrari, Lacasce and
  Merrifield}]{klocker2012reconciling}
Klocker, A., Ferrari, R., Lacasce, J.~H. and Merrifield, S.~T. (2012)
  Reconciling float-based and tracer-based estimates of lateral diffusivities.
\newblock \textit{J. Mar. Res.}, \textbf{70}, 569--602.

\bibitem[{Kunze(1985)}]{kunze85-jpo}
Kunze, E. (1985) Near-inertial wave propagation in geostrophic shear.
\newblock \textit{J. Phys. Oceanogr.}, \textbf{15}, 544--565.

\bibitem[{LaCasce(2008)}]{lacasce2008statistics}
LaCasce, J.~H. (2008) Statistics from {L}agrangian observations.
\newblock \textit{Prog. Oceanogr.}, \textbf{77}, 1--29.

\bibitem[{Lilly et~al.(2011)Lilly, Scott and Olhede}]{lilly11-grl}
Lilly, J.~M., Scott, R.~K. and Olhede, S.~C. (2011) Extracting waves and
  vortices from {L}agrangian trajectories.
\newblock \textit{Geophys. Res. Lett.}, \textbf{38}, 1--5.

\bibitem[{Lumpkin and Pazos(2007)}]{lumpkin07}
Lumpkin, R. and Pazos, M. (2007) Measuring surface currents with {S}urface
  {V}elocity {P}rogram drifters: the instrument, its data, and some results.
\newblock In \textit{{L}agrangian analysis and prediction of coastal and ocean
  dynamics} (eds. A.~Griffa, A.~D. Kirwan, A.~J. Mariano, T.~{\"O}zg{\"o}kmen
  and T.~Rossby), chap.~2, 39--67. Cambridge University Press.

\bibitem[{Mandelbrot and Van~Ness(1968)}]{mandelbrot1968fractional}
Mandelbrot, B.~B. and Van~Ness, J.~W. (1968) Fractional {B}rownian motions,
  fractional noises and applications.
\newblock \textit{SIAM rev.}, \textbf{10}, 422--437.

\bibitem[{McClean et~al.(2008)McClean, Jayne, Maltrud and
  Ivanova}]{mcclean08-omer}
McClean, J., Jayne, S.~R., Maltrud, M.~E. and Ivanova, D. (2008) The fidelity
  of ocean models with explicit eddies.
\newblock In \textit{Ocean modeling in an eddying regime} (eds. M.~W. Hecht and
  H.~Hasumi), 149--164. American Geophysical Union.

\bibitem[{Melard and Schutter(1989)}]{melard1989contributions}
Melard, G. and Schutter, A.~H. (1989) Contributions to evolutionary spectral
  theory.
\newblock \textit{J. Time Series Analysis}, \textbf{10}, 41--63.

\bibitem[{Percival and Walden(1993)}]{percival1993spectral}
Percival, D.~B. and Walden, A.~T. (1993) \textit{Spectral analysis for physical
  applications: {M}ultitaper and conventional univariate techniques}.
\newblock Cambridge University Press.

\bibitem[{Percival and Walden(2000)}]{percival2006wavelet}
--- (2000) \textit{Wavelet methods for time series analysis}.
\newblock Cambridge University Press.

\bibitem[{Pollard and Millard(1970)}]{pollard1970comparison}
Pollard, R.~T. and Millard, R.~C. (1970) Comparison between observed and
  simulated wind-generated inertial oscillations.
\newblock \textit{Deep-Sea Res.}, \textbf{17}, 813--821.

\bibitem[{Priestley(1965)}]{priestley1965evolutionary}
Priestley, M.~B. (1965) Evolutionary spectra and non-stationary processes.
\newblock \textit{J. R. Statist. Soc. B}, \textbf{27}, 204--237.

\bibitem[{Rhines(1979)}]{rhines79-arfm}
Rhines, P.~B. (1979) Geostrophic turbulence.
\newblock \textit{Annu. Rev. Fluid Mech.}, \textbf{11}, 404--441.

\bibitem[{Robinson(1989)}]{robinson89}
Robinson, P.~M. (1989) Nonparametric estimation of time-varying parameters.
\newblock In \textit{Statistical Analysis and Forecasting of Economic
  Structural Change} (ed. P.~Hackl), 253--264. Springer.

\bibitem[{Robinson(1995)}]{robinson1995gaussian}
--- (1995) Gaussian semiparametric estimation of long range dependence.
\newblock \textit{Ann. Statist.}, \textbf{23}, 1630--1661.

\bibitem[{Rossby(2007)}]{rossby07-lapcod}
Rossby, T. (2007) Evolution of {L}agrangian methods in oceanography.
\newblock In \textit{{L}agrangian analysis and prediction in coastal and ocean
  processes} (eds. A.~Griffa, A.~D. Kirwan, A.~J. Mariano, T.~{\"O}zg{\"o}kmen
  and T.~Rossby), chap.~1, 1--38. Cambridge University Press.

\bibitem[{Roueff and Von~Sachs(2011)}]{roueff2011locally}
Roueff, F. and Von~Sachs, R. (2011) Locally stationary long memory estimation.
\newblock \textit{Stoch. Proc. Appl.}, \textbf{121}, 813--844.

\bibitem[{Rupolo et~al.(1996)Rupolo, Hua, Provenzale and
  Artale}]{rupolo1996lagrangian}
Rupolo, V., Hua, B.~L., Provenzale, A. and Artale, V. (1996) Lagrangian
  velocity spectra at 700m in the western {N}orth {A}tlantic.
\newblock \textit{J. Phys. Oceanogr.}, \textbf{26}, 1591--1607.

\bibitem[{Sanderson and Booth(1991)}]{sanderson1991fractal}
Sanderson, B.~G. and Booth, D.~A. (1991) The fractal dimension of drifter
  trajectories and estimates of horizontal eddy-diffusivity.
\newblock \textit{Tellus A}, \textbf{43}, 334--349.

\bibitem[{Sawford(1991)}]{sawford1991reynolds}
Sawford, B.~L. (1991) Reynolds number effects in {L}agrangian stochastic models
  of turbulent dispersion.
\newblock \textit{Phys. Fluids A-Fluid}, \textbf{3}, 1577.

\bibitem[{Schofield et~al.(2007)Schofield, Bishop, MacLean, Brown, Baker,
  Katselidis, Dimopoulos, Pantis and Hays}]{schofield2007novel}
Schofield, G., Bishop, C.~M., MacLean, G., Brown, P., Baker, M., Katselidis,
  K.~A., Dimopoulos, P., Pantis, J.~D. and Hays, G.~C. (2007) Novel {GPS}
  tracking of sea turtles as a tool for conservation management.
\newblock \textit{J. Exp. Mar. Biol. Ecol.}, \textbf{347}, 58--68.

\bibitem[{Stein(1999)}]{stein1999interpolation}
Stein, M.~L. (1999) \textit{Interpolation of spatial data: some theory for
  kriging}.
\newblock Springer Verlag.

\bibitem[{Steinhaeuser et~al.(2012)Steinhaeuser, Ganguly and
  Chawla}]{steinhaeuser2012multivariate}
Steinhaeuser, K., Ganguly, A.~R. and Chawla, N.~V. (2012) Multivariate and
  multiscale dependence in the global climate system revealed through complex
  networks.
\newblock \textit{Clim. Dynam.}, \textbf{39}, 889--895.

\bibitem[{Sykulski et~al.(2015)Sykulski, Olhede, Lilly and
  Early}]{sykulski2013whittle}
Sykulski, A.~M., Olhede, S.~C., Lilly, J.~M. and Early, J.~J. (2015) On
  parametric modelling and inference for complex-valued time series.
\newblock \textit{arXiv preprint arXiv:1306.5993}.

\bibitem[{Thomas(2007)}]{thomas07-aha}
Thomas, L.~N. (2007) Dynamical constraints on the extreme low values of the
  potential vorticity in the ocean.
\newblock In \textit{{Proceedings of the 15th 'Aha Huliko'a Hawaiian Winter
  Workshop on Extreme Events}}. University of Hawaii.

\bibitem[{Thomson(1982)}]{thomson82-ieee}
Thomson, D.~J. (1982) Spectrum estimation and harmonic analysis.
\newblock \textit{Proc. IEEE}, \textbf{70}, 1055--1096.

\bibitem[{Van~Bellegem and Dahlhaus(2006)}]{van2006semiparametric}
Van~Bellegem, S. and Dahlhaus, R. (2006) Semiparametric estimation by model
  selection for locally stationary processes.
\newblock \textit{J. R. Statist. Soc. B}, \textbf{68}, 721--746.

\bibitem[{Veneziani et~al.(2004)Veneziani, Griffa, Reynolds and
  Mariano}]{veneziani2004oceanic}
Veneziani, M., Griffa, A., Reynolds, A.~M. and Mariano, A.~J. (2004) Oceanic
  turbulence and stochastic models from subsurface {L}agrangian data for the
  northwest {A}tlantic {O}cean.
\newblock \textit{J. Phys. Oceanogr.}, \textbf{34}, 1884--1906.

\end{thebibliography}
\end{document}